\documentclass[twocolappendix]{emulateapj}
\usepackage{subfigure}
\usepackage{amsmath}

\newcommand{\bmath}[1]{\mbox{\boldmath{$#1$}}}

\newcommand{\del}{{\bf \nabla}}
\newcommand{\alf}{{\rm Alfv\acute{e}n}}
\newcommand{\cs}{c_{\rm s}}
\newcommand{\va}{v_{\rm A}}

\begin{document}

\title{Signatures of MRI-Driven Turbulence in Protoplanetary Disks: \\ Predictions for ALMA Observations}

\author{Jacob B. Simon\altaffilmark{1,2,3}, A. Meredith Hughes\altaffilmark{4}, Kevin M. Flaherty\altaffilmark{4}, Xue-Ning Bai\altaffilmark{5,6}, 
and Philip J. Armitage\altaffilmark{3,7}}

\email{jbsimon.astro@gmail.com}

\begin{abstract}
Spatially resolved observations of molecular line emission have the potential to yield unique constraints on the nature of turbulence within protoplanetary disks. Using a combination of local non-ideal magnetohydrodynamic simulations and radiative transfer calculations, tailored to properties of the disk around HD 163296, we assess the ability of ALMA to detect turbulence driven by the magnetorotational instability (MRI). Our local simulations show that the MRI produces small-scale turbulent velocity fluctuations that increase in strength with height above the mid-plane. For a set of simulations at different disk radii, we fit a Maxell-Boltzmann distribution to the turbulent velocity and construct a turbulent broadening parameter as a function of radius and height. We input this broadening into radiative transfer calculations to quantify observational signatures of MRI-driven disk turbulence. We find that the ratio of the peak line flux to the flux at line center is a robust diagnostic of turbulence that is only mildly degenerate with systematic uncertainties in disk temperature. For the CO(3-2) line, which we expect to probe the most magnetically active slice of the disk column, variations in the predicted peak-to-trough ratio between our most and least turbulent models span a range of approximately 15\%. Additional independent constraints can be derived from the morphology of spatially resolved line profiles, and we estimate the resolution required to detect turbulence on different spatial scales. We discuss the role of lower optical depth molecular tracers, which trace regions closer to the disk mid-plane where velocities in MRI-driven models are systematically lower.
\end{abstract} 


\altaffiltext{1}{Department of Space Studies, Southwest Research Institute, Boulder, CO 80302}
\altaffiltext{2}{Sagan Fellow}
\altaffiltext{3}{JILA, University of Colorado and NIST, 440 UCB, Boulder, CO 80309-0440}
\altaffiltext{4}{Astronomy Department, Van Vleck Observatory, Wesleyan University, 96 Foss Hill Dr., Middletown, CT 06459}
\altaffiltext{5}{Hubble Fellow}
\altaffiltext{6}{Harvard-Smithsonian Center for Astrophysics, 60 Garden St.,
MS-51, Cambridge, MA 02138}
\altaffiltext{7}{Department of Astrophysical and Planetary Sciences, University of Colorado, Boulder, CO 80309}

\section{Introduction} 

Turbulence is the leading candidate for transporting angular momentum in gaseous 
accretion disks \citep{shakura73}, thus driving evolution of the disk surface density \citep{lynden-bell74}. 
In protoplanetary disks, turbulence is also important for planet formation processes due to coupling between 
gas-phase turbulence and solid bodies. For small, aerodynamically coupled particles, turbulence can lead 
to radial and vertical diffusion \citep{youdin07,clarke88}, while simultaneously 
concentrating particles between vortices on small scales \citep{cuzzi08}, and within pressure maxima 
created by vortices \citep {barge95} or zonal flows \citep{johansen09a} on large scales.
For planets, whose coupling to the gas disk is gravitational rather than aerodynamic, 
turbulence can affect migration via the generation of turbulent torques \citep{nelson04,lubow11} and 
its effect on the saturation of co-orbital resonances \citep{baruteau11,paardekooper11}.

Despite its fundamental importance, observational constraints on the nature of disk turbulence are 
limited and typically indirect. In protoplanetary disks, estimates of disk lifetimes \citep{haisch01} and measurement 
of age-dependent stellar accretion rates \citep{gullbring98,hartmann98a} can be interpreted as requiring 
$\alpha \sim 0.01$ \citep[adopting the $\alpha$ prescription for turbulent angular momentum transport;][]{shakura73}. The best prospects for 
improving upon this rough estimate lie in the detection of non-thermal turbulent broadening of molecular lines, 
either in the infrared \citep{carr04} or the submillimeter \citep{hughes11,guilloteau12,degregorio-monsalvo13}. These measurements are challenging. 
Very roughly, we expect velocity fluctuations to be of the order of $\delta v \sim \alpha^{1/2} \cs \sim 0.1 \cs$, where the 
sound speed $\cs$ is itself smaller than the orbital velocity $v_K$ by a factor of the geometric thickness $(h/r)~\sim~0.05$-$0.15$.  
Discerning the impact of turbulence on line profiles thus requires modeling of the kinematic and thermal structure 
of the disk at better than the 1\% and 10\% level, respectively. This is now feasible for a subset 
of well-resolved disk systems with accurately measured Keplerian rotation profiles. \cite{hughes11} constrained turbulent velocity fluctuations in the upper layers of the disks surrounding TW Hydra and HD~163296 using the CO (3-2) line.  For TW Hydra, they found an upper limit of $\delta v < 0.1 \cs$ and in HD 163296 they measured $\delta v \sim 0.4 \cs$.  Using CS lines, \cite{guilloteau12} determined that $\delta v \sim 0.4-0.5 \cs$ in the molecular layer of the disk around DM Tau. 
Their measurements further suggested that this velocity is independent of height above the mid-plane.

Observational determination of $\delta v$ yields direct insight into the {\it magnitude} of turbulence, but does not provide 
a model-independent constraint on angular momentum transport, particle concentration, or solid body migration. 
The importance of turbulence in these processes depends upon the {\em nature} of the turbulence as well 
as its strength. Convection, for example, is a source of turbulence that is 
expected to be inefficient at transport \citep[e.g.,][]{stone96,lesur10a}, while in the opposite direction, large-scale internal magnetic stresses or 
disk winds \citep[e.g.,][]{salmeron07,suzuki09,bai13a,lesur13,fromang13} can 
transport or extract angular momentum from a near-laminar disk. Even if we exclude winds, 
there are multiple possible drivers of gas-phase turbulence in protoplanetary disks, including 
 gravitational instability \citep{toomre64}, the vertical shear instability \citep{nelson13}, 
the subcritical baroclinic instability \citep{lesur10b}, and the magnetorotational instability \cite[MRI;][]{balbus98}. 
Discriminating between these alternatives may, however, be possible by comparing theoretical predictions 
for how $\delta v$ varies with radius and height in the disk against observational data.

In this paper, we develop theoretical predictions for the radial and vertical variation of molecular 
line profiles from a disk in which angular momentum transport is predominantly driven by the MRI. Preliminary 
work by \cite{simon11b} has shown that the MRI results in turbulent velocity fluctuations $\delta v \sim 0.1 \cs$ at high disk altitudes, 
and that the turbulent velocity is an increasing  
function of height above the mid-plane, particularly within the Ohmic dead zone \citep{gammie96}. This is consistent with previous results \citep{fromang06b}. The trend in MRI-driven turbulence is distinctly different from that measured in simulations of gravitationally unstable disks, in which $\delta v$ is almost independent of height \citep{forgan12,shi14}.

To make useful predictions for future observations requires substantial improvements over these prior calculations. 
First, we need a more accurate treatment of non-ideal magnetohydrodynamic (MHD) effects, which are central to the  
outcome of the MRI in protoplanetary disks \citep{armitage11} due to the low ionization fraction in these systems \citep{wardle07,bai11b}. At 
high densities, Ohmic diffusion can prevent coupling of the magnetic field to the disk gas, leading to the dead zone paradigm 
of \cite{gammie96} in which MRI-turbulent surface layers are ionized due to X-rays and cosmic rays and surround a low-ionization 
dead zone in which there is only very weak accretion \citep{fleming03}. At low gas densities, ambipolar diffusion reduces the 
efficiency of the MRI due to weak collisional coupling between ions and neutrals, leading to the ambipolar damping zone \citep{bai11a,mohanty13,simon13a,simon13b} in the outer disk, and quenched turbulence in the low density disk atmosphere \citep{bai13b,simon13a,simon13b}. At intermediate densities, the Hall effect leads to a qualitatively different behavior in which the magnitude and nature of magnetohydrodynamic transport depends on the orientation of the net vertical field with respect to the disk angular momentum vector \citep{kunz13,lesur14,bai14a}. Here, 
we calculate a series of local disk models, incorporating both Ohmic and ambipolar diffusion, whose physical properties 
(density, temperature, accretion rate) are tailored to match specific radii in the disk around HD~163296 \cite[at the radii of greatest interest 
observationally, we believe that the Hall effect is sub-dominant; e.g.,][Simon 2015, in prep]{bai14b}. Second, we need to derive the directly observable 
quantities (molecular line profiles and channel maps) from the primary theoretical output (velocity fields at different levels within the disk). 
To carry out this task, we use the {\sc LIME} radiative transfer code in conjunction with an empirical model for the vertical thermal 
structure of the disk to generate molecular lines and channel maps. This approach
allows us to make the first realistic predictions for sub-mm observations of HD~163296 under 
the assumption that any turbulence in the system derives from the MRI. These predictions will be testable with 
forthcoming high resolution, high sensitivity ALMA observations of the HD~163296 disk.

The outline of this paper is as follows.  In Section~\ref{model}, we describe the empirical model for the HD~163296 disk that we use throughout this work. In Section~\ref{simulations}, we first explain our numerical
methodology and then discuss the results from our simulations, including the extraction of the turbulent velocity for input into the {\sc LIME} code.  Our radiative transfer calculations and observational predictions are presented
in Section~\ref{predictions}, and we summarize and present our conclusions in Section~\ref{summary}.

\section{Disk Model}
\label{model}

Our numerical simulations and radiative transfer calculations are based on a representative model for the disk around HD~163296, 
which is broadly consistent with previous observations of this source \citep{rosenfeld13}. The disk model is needed for two distinct 
purposes in this work. First, the surface density and mid-plane temperature (which sets the vertical scale height) at a small number of 
discrete radii serve as input to 
the local MHD simulations that we use to determine the predicted turbulent velocity. The simulations are isothermal, 
and hence are characterized by a single temperature and have an approximately Gaussian vertical density profile (modified only by 
the effects of magnetic pressure). Second, an axisymmetric disk model is needed as input to the radiative transfer calculations that we 
use to determine the predicted line shapes and channel maps. A vertically isothermal disk structure would yield results that are grossly 
inconsistent with observations because we are interested in line emission that originates from near the surface of the disk where the 
temperature is substantially elevated. Accordingly, for the radiative transfer calculations we use a version of the disk model in which 
$T$ is a function of height as well as radius, though we continue to assume a Gaussian density profile. This approach is computationally 
expedient -- non-isothermal disk simulations are significantly more difficult to develop and run -- but it carries a cost in terms of 
consistency; neither the radiative transfer model nor the matching between the simulations and the radiative transfer can be fully 
self-consistent.

The vertical temperature structure is consistent with warm surface layers surrounding a cooler mid-plane. We first define the ``mid-plane" and ``surface layer" temperatures, respectively, as

\begin{equation}
\label{tm}
T_{\rm m}(r) = T_{\rm m,0} \left(\frac{r}{155{\rm AU}}\right)^{-0.3}
\end{equation}

\begin{equation}
T_{\rm s}(r) = T_{\rm s,0} \left(\frac{r}{200{\rm AU}}\right)^{-0.5}.
\end{equation}

\noindent
The complete gas temperature is 

\begin{equation}
\footnotesize
T(r,z) = \left\{ \begin{array}{ll}
T(r,z) = T_{\rm s}(r) & \quad  \mbox{$|z| \ge z_q$} \\
T(r,z) = T_{\rm s}(r) + \left[T_{\rm m}(r) - T_{\rm s}(r)\right] \left[{\rm cos}\left(\frac{\pi z}{2 z_q}\right)\right]^{2\delta} & \quad  \mbox{$|z| < z_q$} 
\end{array}\right.
\end{equation}

\noindent
where $z_q$ and $\delta$ depend on the radius $r$ via,

\begin{equation}
z_q = 63{\rm AU} \left(\frac{r}{200{\rm AU}}\right)^{1.3}{\rm Exp}\left[-\left(\frac{r}{800{\rm AU}}\right)^2\right]
\end{equation}

\noindent
and

\begin{equation}
\delta = 0.0034 \left[\left(\frac{r}{\rm 1AU}\right)-200\right]+2.5 .
\end{equation}

The surface density for HD~163296 is well fit by a power law with an exponential drop off at large radii:

\begin{equation}
\Sigma(r) = M_{\rm disk} \left(\frac{2-\gamma_D}{2\pi R_c^2}\right) \left(\frac{r}{R_c}\right)^{-\gamma_D} {\exp}\left[-\left(\frac{r}{R_c}\right)^{2-\gamma_D}\right] 
\end{equation}

\noindent
from which we can calculate the volume mass density,

\begin{equation}
\rho(r,z) = \frac{\Sigma(r)}{\sqrt{\pi}H(r)}{\exp}\left[-\left(\frac{z}{H(r)}\right)^2\right]
\end{equation}

\noindent
where the vertical scale height $H(r)$ is

\begin{equation}
\label{scale_height}
H = \frac{\sqrt{2} c_s}{\Omega}.
\end{equation}

\noindent
The sound speed (which we take to be isothermal here given the Gaussian density profile) is

\begin{equation}
\label{sound_speed}
c_s = \sqrt{\frac{{k_{\rm boltz}} T_{\rm m}(r)}{\mu m_{\rm H}}}
\end{equation}

\noindent
where $T_{\rm m}$ is the mid-plane temperature profile from Equation~(\ref{tm}), and 

\begin{equation}
\Omega = \sqrt{\frac{G M_{\rm star}}{r^3}}
\end{equation}

\noindent
appropriate for Keplerian rotation.  ${k_{\rm boltz}}$ is Boltzmann's constant, $m_{\rm H}$ is the mass of hydrogen, and $G$ is the gravitational constant.

Finally, our radiative transfer model includes the abundance of CO relative to hydrogen, $X_{\rm CO}$, in order to produce synthetic observables using CO lines.   In most regions, we assume $X_{\rm CO} = 10^{-4}$. 
If the temperature is less than 19K, we set $X_{\rm CO} = 10^{-12}$, representing freeze-out of CO onto dust grains.  Photo-dissociation also depletes CO in the surface layers. 
When $N < 5\times10^{20}~{\rm cm}^{-2}$ (where $N$ is the integrated number density from the surfaces of the disk), 
 we set $X_{\rm CO} = 10^{-12}$.  We can translate this to a condition on the complementary error function of $z$ because the gas density is a simple Gaussian in $z$.  Thus, after some basic arithmetic, we set the CO abundance to this low value if Erfc$[z/H(r)] < 10^{17} m_{\rm H}/\Sigma(r)$ and if Erfc$[-z/H(r)] < 10^{17} m_{\rm H}/\Sigma(r)$ (here, the pre factor is assumed to have appropriate units to make the ratio dimensionless).

Our model disk parameters are presented in Table~\ref{tbl:model_parameters}.

\begin{deluxetable}{c|c}
\tabletypesize{\scriptsize}
\tablewidth{0pc}
\tablecaption{Model Parameters\label{tbl:model_parameters}}
\tablehead{
\colhead{Parameter}&
\colhead{Value}    }
\startdata
$T_{\rm m,0}$ & 19 K \\
$T_{\rm s,0}$ & 40 K \\
$M_{\rm disk}$ & 0.09 $M_{\sun}$ \\
$M_{\rm star}$ & 2.3 $M_{\sun}$ \\
$\gamma_D$ & 0.8 \\
$R_c$ & 115 AU \\
$\mu$ & 2.37 \\
\enddata
\end{deluxetable}

\section{Local Disk Simulations}
\label{simulations}

Our numerical simulations are a series of local, disk simulations of an accretion disk patch placed at several different radii in the model for HD~163296 described above. 
Here, we describe the numerical algorithms and initial conditions for these simulations.
In this paper, there will be two sets of units.  The first set of units will be {\it code units}, which we will refer to when we are talking about specifics of numerical simulations.  The second set of units will be {\it physical units}, which will apply to our radiative transfer calculations and all other analysis from that point on.

\subsection{Numerical Method and Setup}
\label{setup}

Our simulations use \textit{Athena}, a second-order accurate Godunov
flux-conservative code for solving the equations of MHD. 
\textit{Athena} uses the dimensionally unsplit corner transport upwind method
of \cite{colella90} coupled with the third-order in space piecewise
parabolic method of \cite{colella84} and a constrained transport
\citep[CT;][]{evans88} algorithm for preserving the $\del \cdot {\bmath
B}$~=~0 constraint.  We use the HLLD Riemann solver to calculate the
numerical fluxes \cite[]{miyoshi05,mignone07b}.  A detailed description
of the base \textit{Athena} algorithm and the results of various test problems
are given in \cite{gardiner05a}, \cite{gardiner08}, and \cite{stone08}.

The simulations employ a local shearing box approximation.
The shearing box models a co-rotating disk patch whose size is small compared to the
radial distance from the central object, $R_0$.  This allows the
construction of a local Cartesian frame $(x,y,z)$ that is defined in terms of the disk's
cylindrical co-ordinates $(R,\phi,z^\prime)$ via  $x=(R-R_0)$, $y=R_0 \phi$, and $z = z^\prime$.
The local patch  co-rotates with an angular velocity $\Omega$ corresponding to
the orbital frequency at $R_0$, the center of the box; see \cite{hawley95a}.  The equations to solve are:

\begin{equation}
\label{continuity_eqn}
\frac{\partial \rho}{\partial t} + \del \cdot (\rho {\bmath v}) = 0,
\end{equation}
\begin{equation}
\label{momentum_eqn}
\begin{split}
\frac{\partial \rho {\bmath v}}{\partial t} + \del \cdot \left(\rho {\bmath v}{\bmath v} - {\bmath B}{\bmath B}\right) + \del \left(P + \frac{1}{2} B^2\right) \\
= 2 q \rho \Omega^2 {\bmath x} - \rho \Omega^2 {\bmath z} - 2 {\bmath \Omega} \times \rho {\bmath v} \\
\end{split}
\end{equation}
\begin{equation}
\label{induction_eqn}
\frac{\partial {\bmath B}}{\partial t} - \del \times \left({\bmath v} \times {\bmath B}\right) = -\del \times \left(\eta_{\rm A}{\bmath J_{\perp}}+\eta_{\rm O} {\bmath J}\right),
\end{equation} 

\noindent 
where $\rho$ is the mass density, $\rho {\bmath v}$ is the momentum
density, ${\bmath B}$ is the magnetic field, $P$ is the gas pressure,
and $q$ is the shear parameter, defined as $q = -d$ln$\Omega/d$ln$R$.
We use $q = 3/2$, appropriate for a Keplerian disk.  For simplicity and numerical convenience, we
assume an isothermal equation of state $P = \rho \cs^2$, where $\cs$
is the isothermal sound speed.  From left to right, the source terms
in equation~(\ref{momentum_eqn}) correspond to radial tidal forces
(gravity and centrifugal), vertical gravity, and the Coriolis force. The first electromotive force (EMF) term on the right-hand-side of equation~(\ref{induction_eqn}) is ambipolar diffusion, which consists of diffusivity $\eta_{\rm A} = B^2/\gamma\rho\rho_i$ multiplied by the component of the current density ${\bmath J}$ perpendicular to the magnetic field, ${\bmath J_{\perp}}$.   Here,  $\rho_i$ is the ion density, and
$\gamma$ is the coefficient of momentum transfer for ion-neutral
collisions.  Ohmic diffusion is included via the second EMF term on the right-hand-side and is the diffusivity $\eta_{\rm O}$ multiplied by ${\bmath J}$.
Our system of code units has the magnetic permeability equal to unity, and
the current density is

\begin{equation}
\label{current}
{\bmath J} = \del \times {\bmath B}.
\end{equation}

Numerical algorithms for integrating these equations are described in detail in
\cite{stone10} (see also the Appendix of \citealp{simon11a}). 
The $y$ boundary conditions are strictly periodic, whereas the $x$ boundaries
are shearing periodic \cite[]{hawley95a}. The vertical boundary conditions are the
modified outflow boundaries described in \cite{simon13b}. The EMFs at
the radial boundaries are properly remapped to guarantee that the net
vertical magnetic flux is conserved to machine precision using CT
\citep{stone10}. 

As in \cite{bai11a}, \cite{simon11b}, \cite{simon13a}, and \cite{simon13b}, both Ohmic and ambipolar diffusion are implemented in a first-order in time operator-split
manner using CT to preserve the divergence free condition
with an additional step of remapping $J_y$ at radial shearing-box boundaries.
The super time-stepping (STS) technique of \cite{alexiades96} has been implemented to
accelerate our calculations (see the Appendix of \cite{simon13a}).  

We calculate the Ohmic and ambipolar diffusivities by interpolating from pre-computed diffusivity tables based on equilibrium chemistry, following the methodology described in \cite{bai13b} and subsequent works \citep{bai13c,bai14a}. The diffusivity tables give $\eta_{\rm O}$  and $\eta_{\rm A}/B^2$ as a function of density and ionization rate at fixed temperature. They are independent of $B$ for the regimes we consider in this paper (in the absence of abundant small grains). The chemical reaction network is described in \cite{bai09} and \cite{bai11b}, and recently updated in \cite{bai14a} with the latest version of the UMIST database \citep{mcelroy13}.  We consider both grain-free chemistry, as well as a chemistry model containing $0.1\mu{\rm m}$ grains with abundance of $10^{-4}$. In the simulations, the ionization rates are obtained based on the horizontally averaged column density to the disk surface, $\Sigma$. We include contributions from stellar X-rays using standard prescriptions \citep{igea99}, with fitting formulas provided by \cite{bai09}, and cosmic rays with ionization rate $\xi_{\rm cr}=10^{-17}\exp{\left[-\Sigma/(96{\rm~g~cm}^{-2})\right]}$s$^{-1}$ \citep{umebayashi81}. For X-ray ionization, we use an X-ray luminosity $L_X= 4\times10^{29}$ erg s$^{-1}$ and X-ray temperature corresponding to 1 keV, specifically appropriate to HD163296 \citep{swartz05,gunther09}. In the surface layer, we further include the effect of far UV ionization based on calculations from \cite{perez-becker11b}. The FUV ionization is assumed to have a constant penetration depth of $\Sigma_{\rm FUV}$ in the range of $0.01-0.1$~g~cm$^{-2}$. 

The degree of FUV ionization is critically important for MHD turbulence in the outer disk, because only FUV photons can yield  
ionization levels in the upper disk layers that are high enough to prevent MRI quenching by ambipolar diffusion. We treat 
the effect of FUV ionization on ambipolar diffusion using 
the same procedure as in \cite{simon13b}. We define the ambipolar diffusion Elsasser number,

\begin{equation}
\label{am1}
{\rm Am}\equiv\frac{\gamma\rho_i}{\Omega},
\end{equation}

\noindent
which corresponds to the number of times a neutral molecule collides with the ions in a
dynamical time ($\Omega^{-1}$). Since in the FUV layer, Am $\gg$ 1 (i.e., ambipolar diffusion is weak) at all radii, we simply adopt the function from \cite{simon13b}, which is sufficient for our purposes.
In the FUV ionization layer, Am is

\begin{equation}
\label{Am_FUV}
{\rm Am_{\rm FUV}} \approx3.3\times10^7
\bigg(\frac{f}{10^{-5}}\bigg)\bigg(\frac{\rho}{\rho_{0,{\rm mid}}}\bigg)\left(\frac{r}{{\rm 1 AU}}\right)^{-5/4}\ ,
\end{equation}

\noindent
The Am value below the FUV ionization layer, Am$_{\rm mid}$ is determined by the diffusivity table, as described above.  Piecing these two regions together, we have the following complete functional form for Am,

\begin{equation}
\label{amc}
\small
{\rm Am} \equiv \left\{ \begin{array}{ll}
 {\rm Am_{\rm FUV}} & \quad 
\mbox{$z \ge z_t + \Delta z$} \\
{\rm Am}_{\rm mid} + \frac{1}{2}{\rm Am_{\rm FUV}}S^+(z)  & \quad
\mbox{$z_t - n\Delta z < z < z_t + \Delta z$} \\
{\rm Am}_{\rm mid} & \quad
 \mbox{$z_b+n\Delta z \le z \le z_t - n \Delta z$} \\
{\rm Am}_{\rm mid} + \frac{1}{2}{\rm Am_{\rm FUV}}S^-(z) & \quad
\mbox{$z_b-\Delta z < z < z_b + n\Delta z$} \\
{\rm Am_{\rm FUV}} & \quad
\mbox{$z \le z_b - \Delta z$}
\end{array} \right.
\end{equation}

\noindent
where $z_t$ and $z_b$ are the top and bottom layers of the FUV ionization layer, respectively, and are calculated by integrating at each time step the horizontally averaged column density
from the boundary towards the mid-plane until $\Sigma_{\rm FUV}$ is reached.  $S^+(z)$ and $S^-(z)$ are smoothing functions defined as 
 
\begin{equation}
\small
\label{splus}
S^+(z) \equiv 1+{\rm Erf}\left(\frac{z-0.9z_t}{\Delta z}\right),
\end{equation}
\begin{equation}
\small
\label{sminus}
S^-(z) \equiv 1-{\rm Erf}\left(\frac{z-0.9z_b}{\Delta z}\right),
\end{equation}

\noindent
Here, $n = 8$ and $\Delta z$ ranges from $0.1 H$ to $0.05 H$, depending on the simulation.  These numbers were chosen to give a reasonably
resolved transition region between Am = Am$_{\rm mid}$  and ${\rm Am_{\rm FUV}}$.

We center the shearing box simulations on several radii in the mid-to-outer regions of HD~163296, at 10~AU, 30~AU, and 100~AU.  
Recent results \cite[e.g.,][]{bai13a,simon13a,simon13b} point to the importance of the strength of the net vertical magnetic field threading the domain as well as the depth to
which FUV photons ionize the upper disk layers.  As such, we choose these parameters to explore in our parameter study.  In particular, we examine initial field strengths, defined
by the gas to magnetic pressure ratio at the mid-plane, $\beta_0$, of $\beta_0 = 10^4$ and $10^5$.  We explore the limiting cases for the column to which FUV  photons penetrate, taken
from the results of \cite{perez-becker11b}: $\Sigma_{\rm FUV} = 0.01$~g~cm$^{-2}$ and $\Sigma_{\rm FUV} = 0.1$~g~cm$^{-2}$.  We also run one case at each radius that includes the effects
of grain chemistry on the diffusivity coefficients, though we don't find a significant difference between this simulation and the corresponding simulations with no grains.

Aside from the initial field strength and details of the diffusion profile that depend on the inclusion of grain chemistry and the depth of FUV ionization,
all simulations start from the same initial conditions.  The gas density is in hydrostatic
equilibrium for an isothermal gas,

\begin{equation}
\label{density_init}
\rho(x,y,z) = \rho_0 {\rm exp}\left(-\frac{z^2}{H^2}\right),
\end{equation}
where $\rho_0 = 1$ is the mid-plane density in code units, and $H$ is the scale height in the disk defined by the mid-plane gas temperature (see Equations~(\ref{scale_height}) and (\ref{sound_speed}))

In code units, the isothermal sound speed, $\cs = 7.07 \times 10^{-4}$, corresponding to an initial value for
the mid-plane gas pressure of $P_0 = 5\times 10^{-7}$.  With $\Omega = 0.001$, the value for
the scale height is $H = 1$.   A density floor is imposed as
too small a density leads to a large $\alf$ speed and a very small
time step. The value for the density floor depends on the simulation but ranges from $10^{-5}$ to $10^{-4}$ of the initial mid-plane value.

In all runs, the initial magnetic field is a net vertical field. It is well known that for purely
vertical field, the MRI sets in from a transient channel flow. For relatively strong net vertical
magnetic flux, the channel flow is so strong as to cause numerical problems and/or disk
disruption in the simulations \citep{miller00}. To circumvent such potential difficulties, we
add a sinusoidally varying vertical field component so that, 

\begin{equation}
\label{initi_b}
B_z = B_0 \left[1+\frac{1}{2}{\rm sin}\left(\frac{2\pi}{L_x}x\right)\right],
\end{equation}

\noindent
where $L_x$ is the domain size in the $x$ dimension, and 

\begin{equation}
\label{bo}
B_0 = \sqrt{\frac{2P_0}{\beta_0}}.
\end{equation}
Here, $B_0$ is the net vertical magnetic field, characterized by $\beta_0$, described above.  With the asymmetry introduced by the extra sinusoidal variation in the vertical field, the strong growth
of channel flows \cite[see][]{hawley95a,miller00} is suppressed at early stages, and the simulation can integrate beyond the
initial transient without numerical problems. All other magnetic field components are initialized to be zero.

To seed the MRI, random perturbations are added to the density and velocity components.  The amplitude of these perturbations are $\delta\rho/\rho_0$= 0.01 and
$\delta v_i = 0.004 \cs$ for $i=x,y,z$. All simulations are carried out at a resolution of 36 zones per $H$ and at a box size of $4H\times8H\times8H$.

To summarize, we have three free parameters: the initial magnetic field strength, the column to which the FUV photons penetrate, and the inclusion of dust grains in our chemistry calculation of the diffusivity profiles. 
The runs are listed in Table~\ref{tbl:sims}.  The label for each run is given by its radial location, field strength, the FUV depth, and the chemistry calculation.  So, for example, run R10-B4-FUV0.01-NG is placed at a radius of 10 AU,
has a magnetic field strength defined by $\beta_0 = 10^4$, has an FUV penetration column of 0.01 g ${\rm cm}^{-2}$, and has no grains ``NG".

\begin{widetext}
\begin{deluxetable*}{l|ccccrcc}
\tabletypesize{\small}
\tablewidth{0pc}
\tablecaption{Shearing Box Simulations\label{tbl:sims}}
\tablehead{
\colhead{Label}&
\colhead{Radius}&
\colhead{$\beta_0$}&
\colhead{$\Sigma_{\rm FUV}$}&
\colhead{Grains?}&
\colhead{$\alpha$}&
\colhead{$\dot{M}$}&
\colhead{Simulation Category} \\
\colhead{ }&
\colhead{(AU)}&
\colhead{ }&
\colhead{(g cm$^{-2}$)}&
\colhead{ }&
\colhead{ }&
\colhead{($M_{\sun}/{\rm yr}$)}&
\colhead{ } } 
\startdata
R10-B4-FUV0.1-NG & 10 & $10^4$ & 0.1 & No & 0.0071 & $2\times10^{-7}$ & Non-dynamo  \\
R10-B5-FUV0.01-NG & 10 & $10^5$ & 0.01 & No & 0.00081 &$2\times10^{-8}$ & Non-dynamo \\
R10-B5-FUV0.1-NG & 10 & $10^5$ & 0.1 & No & 0.0024 & $2\times10^{-8}$ & Dynamo \\
R10-B5-FUV0.01-G & 10 & $10^5$ & 0.01 & Yes & 0.00078 & $2\times10^{-8}$ & Non-dynamo \\
\vspace{0.005in} \\
\hline
\vspace{0.005in} \\
R30-B4-FUV0.01-NG & 30 & $10^4$ & 0.01 & No & 0.0052 & $1\times10^{-7}$ & Non-dynamo \\
R30-B4-FUV0.1-NG & 30 & $10^4$ & 0.1 & No & 0.0099 & $2\times10^{-7}$  & Non-dynamo \\
R30-B5-FUV0.01-NG & 30 & $10^5$ & 0.01 & No & 0.0013 & $2\times10^{-8}$ & Non-dynamo \\
R30-B5-FUV0.1-NG & 30 & $10^5$ & 0.1 & No & 0.0032 & $2\times10^{-8}$ & Dynamo \\
R30-B5-FUV0.01-G & 30 & $10^5$ & 0.01 & Yes & 0.0013 &$2\times10^{-8}$ & Non-dynamo \\
\vspace{0.005in} \\
\hline
\vspace{0.005in} \\
R100-B4-FUV0.01-NG & 100 & $10^4$ & 0.01 & No & 0.0075 & $9\times10^{-8}$ & Non-dynamo \\
R100-B4-FUV0.1-NG & 100 & $10^4$ & 0.1 & No & 0.027 & $1\times10^{-7}$ & Dynamo \\
R100-B5-FUV0.01-NG & 100 & $10^5$ & 0.01 & No & 0.0026 & $9\times10^{-9}$ & Dynamo \\
R100-B5-FUV0.1-NG & 100 & $10^5$ & 0.1 & No & 0.0044 & $1\times10^{-8}$ & Dynamo \\
R100-B5-FUV0.01-G & 100 & $10^5$ & 0.01 & Yes & 0.0025 & $8\times10^{-9}$ & Dynamo \\
\enddata
\end{deluxetable*}
\end{widetext}

\subsection{Simulation Results}
\label{simulation_results}

In this section, we first discuss the basic properties of our simulations, and how they compare to existing results in the literature. We then describe our 
methodology for extracting a mean vertical profile of turbulent velocity from our runs for subsequent use in the radiative transfer calculation of molecular line profiles and other observables.

\begin{figure*}[t]
\begin{center}
\includegraphics[width=\textwidth,angle=0]{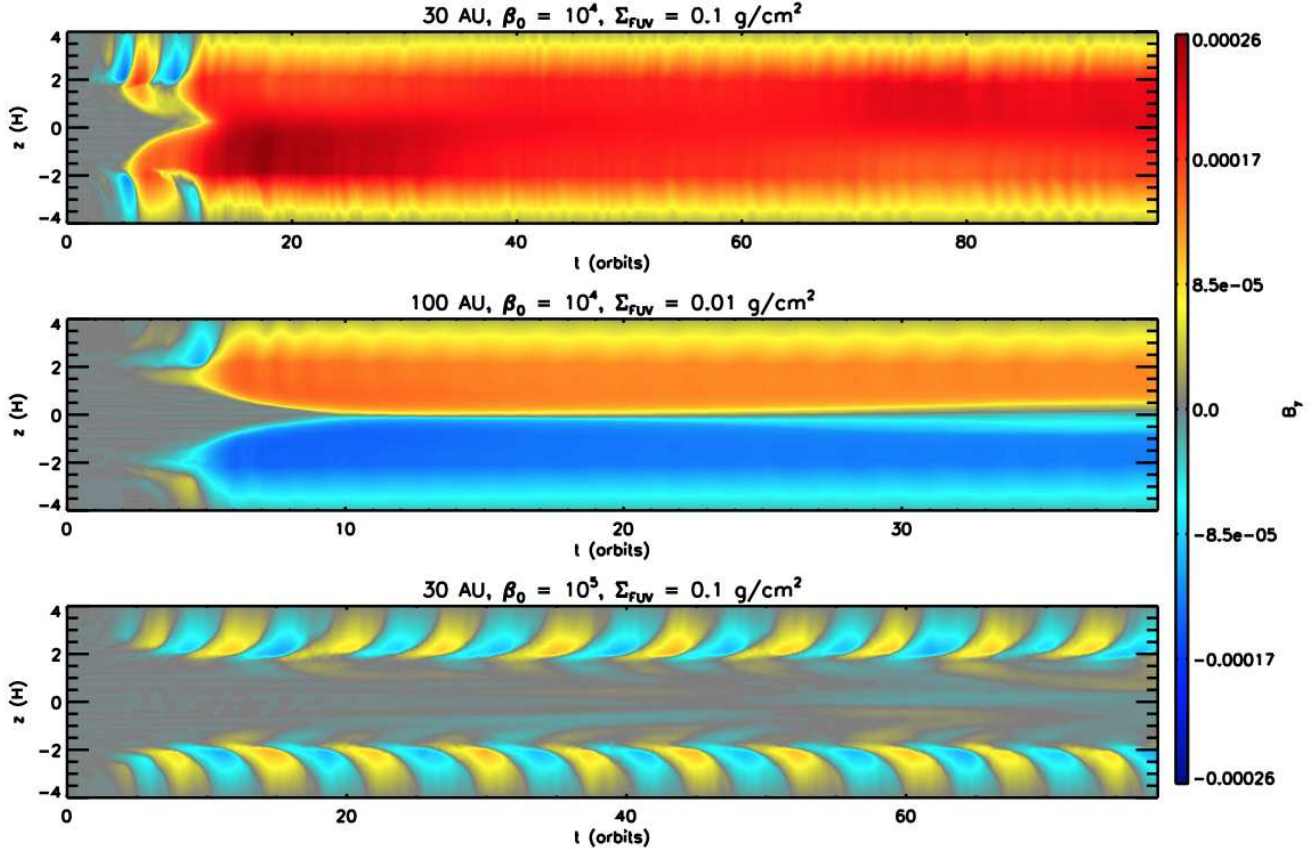}
\end{center}
\caption{
Space-time diagram of the horizontally averaged toroidal field for two non-dynamo simulations (top and middle) and one dynamo simulation (bottom).  For the non-dynamo runs, the toroidal field is largely constant in time at late times and either flips sign near the mid-plane (middle panel) or is roughly constant at all $z$ (top panel).  The dynamo simulation shows MRI dynamo oscillations for $|z| \gtrsim 2H$ and a less active mid-plane region.  These behaviors are representative of our simulations.
}
\label{sttz_by}
\end{figure*}
 
\begin{figure*}[t]
\begin{center}
\includegraphics[width=\textwidth,angle=0]{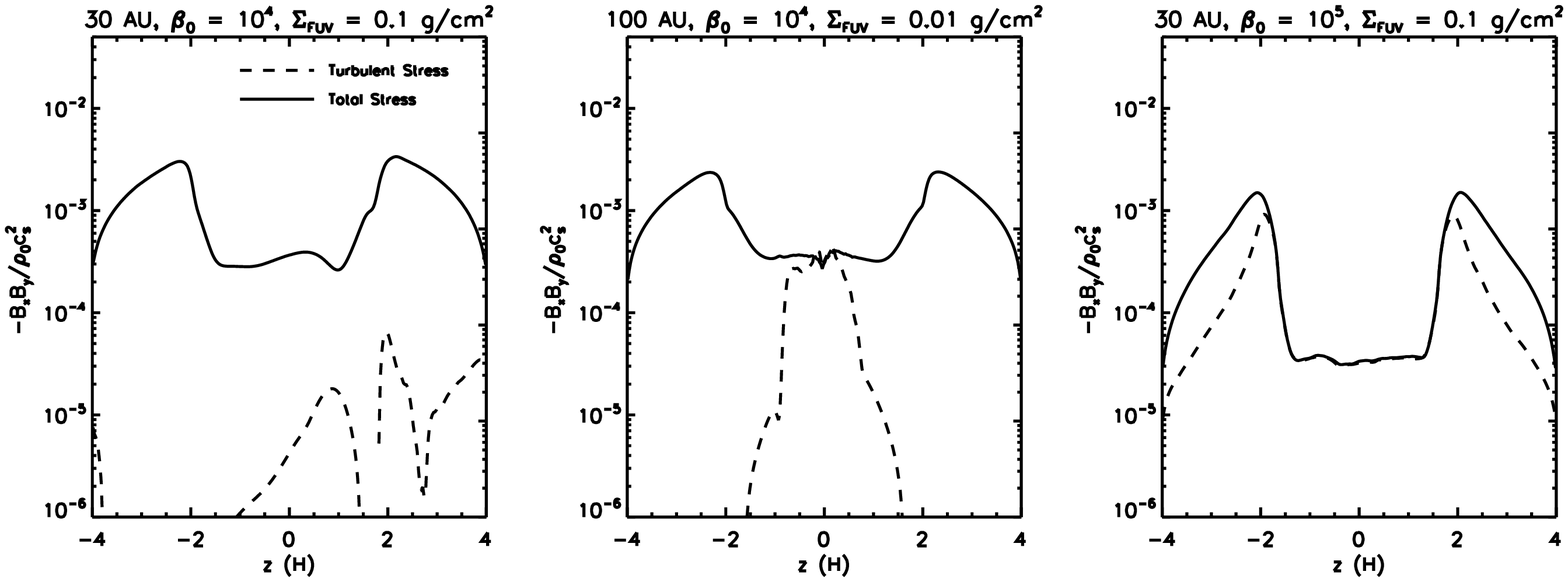}
\end{center}
\caption{
Time and horizontally averaged vertical profile of the $xy$ component of the Maxwell stress for the three representative simulations of Fig.~\ref{sttz_by}. The solid lines are $\overline{\langle-B_xB_y\rangle}$, where the brackets denote a horizontal average and the bar denotes a time average. The dashed lines are the Maxwell stress due to turbulent fluctuations, which is calculated by subtracting the correlation between large scale fields, $-\overline{\langle B_x\rangle\langle B_y\rangle}$, from $\overline{\langle-B_xB_y\rangle}$.   The left panel shows a non-dynamo simulation in which the Maxwell stress is essentially entirely large scale.  The middle panel shows a non-dynamo solution in which the Maxwell stress is small scale near the mid-plane (in this case due to magnetic reconnection of a current sheet) but is otherwise large scale. The right panel shows a dynamo simulation in which there is significant small scale stress due to turbulence at both large $|z|$ (though, large scale stress still dominates here) and in the mid-plane region. 
}
\label{maxwell}
\end{figure*}

\subsubsection{Preliminaries}
\label{preliminaries} 

To set the stage for our turbulent velocity characterization below, we first examine our simulations through several standard diagnostics.  The first 
is  the $\alpha$ parameter of \cite{shakura73}, calculated as the time average of the density weighted $R\phi$ ($xy$ in shearing box coordinates) component of the stress tensor,

\begin{equation}
\label{wrp}
\alpha \equiv \overline{\left[\frac{\left\langle \rho v_x \delta v_y
- B_xB_y\right\rangle}{\left\langle \rho\cs^2\right\rangle}\right]},
\end{equation}

\noindent
where the angled brackets denote a volume average and the over-bar denotes a time average.  The value of $\alpha$ for each simulation is given in Table~\ref{tbl:sims} and generally lies within the range $\sim 0.001-0.01$, consistent with previous shearing box studies of the MRI \cite[e.g.,][]{sano04,simon12}.   The $\alpha$ values also agree with the simulations of \cite{simon13b}, which explored the nature of MRI turbulence in the presence of ambipolar diffusion and a net vertical field in the outer regions of a disk with a Minimum Mass Solar Nebula density profile.
  
The simulations can be classified into two categories based on the evolution and structure of the magnetic field.  The first type of solution, which we refer to as a ``dynamo'' solution, shows the standard MRI dynamo behavior for the magnetic field.  In particular, the mean radial and toroidal fields flip sign with a period of $\sim$10 orbits in the active regions that surround the Ohmic and/or ambipolar diffusion dominated mid-plane region.  The second type of solution, which we call ``non-dynamo", consists of a mean magnetic field that is approximately constant in time.   

Figure~\ref{sttz_by} shows the space-time evolution of the horizontally averaged toroidal field for two non-dynamo simulations (top two panels) and one dynamo simulation (bottom panel).  In previous work \citep{simon13b} we described the non-dynamo simulations as being ``quasi-laminar", but here 
we choose a different naming convention to emphasize that these solutions are not necessarily laminar in nature. In particular, for most of our non-dynamo simulations, there is a significant Maxwell stress contribution from scales other than the largest scale in the box.   This is shown in  Figure~\ref{maxwell}, which shows the time and horizontally averaged vertical profile for the $xy$ component of the Maxwell stress, $\overline{\langle-B_xB_y\rangle}$, for three separate simulations.  Also plotted in the figure is the {\it small scale} contribution to the Maxwell stress, calculated by subtracting the Maxwell stress at the largest scale, $-\overline{\langle B_x\rangle\langle B_y\rangle}$, from $\overline{\langle-B_xB_y\rangle}$.  The degree to which these stresses overlap represents how much of the stress resides at large scales as opposed to turbulent fluctuations.  The left and middle panel show the two non-dynamo simulations from Fig.~\ref{sttz_by}, whereas the panel on the right shows the dynamo simulation.  In the non-dynamo cases, the Maxwell stress can be largely dominated by large scale stress (as in the left panel), but in general these simulations have some small scale turbulence towards the mid-plane region (middle panel).   In the dynamo simulation, the Maxwell stress is generally large scale at large $|z|$, but has a much greater contribution from small scale stress compared to the non-dynamo simulations.  Furthermore, the mid-plane regions of the dynamo simulations are always dominated by small scale turbulent stress. The simulation category is displayed in Table~\ref{tbl:sims}.

One feature of note is the strong asymmetry in the small-scale turbulent stress of the left panel of Fig.~\ref{maxwell}.
We are not entirely sure why this asymmetry exists, but we believe that it results from a combination of
stochastic variation in the magnetic field strength, which generally plays a role in driving small scale stress, and relatively short integration times.  In particular, as
shown in Fig.~\ref{sttz_by} (top panel), there are top-bottom asymmetries in the magnetic field strength that stochastically
appear and disappear.  This stochastic behavior has also been seen in several of the other non-dynamo cases.
The origin of this behavior remains unclear, but it is likely that the asymmetry observed in Fig.~\ref{maxwell}
results from time-averaging over a small number of variations in the magnetic field strength; averaging over many such variations
would likely remove the asymmetry or reduce it substantially.

The dynamo solution becomes more robust at larger radii, where the vertical depth (in terms of physical distance) to which the FUV photons penetrate becomes larger.  As a greater fraction of the vertical extent of the disk is ionized due to FUV photons, regions of lower $\alf$ speed overlap with high values of Am, allowing for the standard MRI dynamo to operate.   Thus, for radii larger than 100 AU, we expect that the disk will be in the dynamo state, at least for the values of $\beta_0$ explored here.

As in the quasi-laminar simulations of \cite{simon13b}, the vertical magnetic field in the non-dynamo simulations can launch a magnetic wind, thus removing angular momentum through a Blandford-Payne type of mechanism \citep{blandford82,lesur13,bai13a,bai13b,fromang13}. The stress corresponding to this angular momentum removal lies in the $zy$ component of the stress tensor and in dimensionless form is given by

\begin{equation}
\label{wzphi}
W_{zy} \equiv \frac{(\rho v_z\delta v_y-B_zB_y)}{\rho_0\cs^2}\bigg|^{z_{\rm top}}_{z_{\rm bot}},
\end{equation}

\noindent
where $\rho_0$ is the initial mid-plane gas density and $z_{\rm top}$ and $z_{\rm bot}$ are the top and bottom of the magnetic wind that is produced as a result of the net vertical field threading the box.   The actual values for $z_{\rm top}$ and $z_{\rm bot}$ are somewhat arbitrary in a shearing box \cite[e.g.,][]{simon13b}, and here, we take them to be the top and bottom of the shearing box; $z_{\rm top} = 4H$ and $z_{\rm bot} = -4H$.   This stress component is present in the dynamo simulations as well, since these simulations also contain a net vertical field.

In the shearing-box approximation, the radially inner and outer sides of the box are symmetric (i.e., curvature is ignored). This leads to two possible types of outflow geometry, depending on whether the outflow from the top and bottom sides of the box flow toward the same (``even-$z$" symmetry) or opposite (``odd-$z$" symmetry) radial directions \citep{bai13b}. The desired (physical) outflow geometry is the ``even-$z$" symmetry. Simulations by \cite{bai13b} and \cite{bai13c} that focused on the inner region of protoplanetary disks ($\lesssim10$ AU) showed that the MRI is completely suppressed due to Ohmic resistivity and ambipolar diffusion, and the even-$z$ symmetry can be achieved with horizontal magnetic field flipped through a thin layer offset from the mid-plane. On the other hand, when the MRI is operating, as in the ideal MHD simulations of \cite{bai13a}, either the odd-$z$ symmetry prevails (when net $B_z$ is sufficiently strong), or the MRI dynamo makes the radial direction of the outflow change sign in time.  We observe this latter feature in our dynamo simulations as well, which argues against the outflow resulting from a physical disk wind. Ultimately, this symmetry issue needs to be resolved with global simulations. Here, we are primarily interested in the turbulent velocity distribution in the disk at heights below those where a wind may be launched, and we need to calculate the $zy$ stress only for the purposes of estimating the accretion rate that our disk model would support at each radius. 
To calculate the $zy$ stress, we assume  that the outflow {\it always} follows the physical ``even-$z$ symmetry" wind discussed above, which gives 

\begin{equation}
\label{wzphi_symm}
W_{zy}\bigg|_{z_{\rm top}} = -W_{zy}\bigg|_{z_{\rm bot}}.
\end{equation}

\noindent
Thus, the quantity of interest for our simulations will be the average of $2 |W_{zy}|_{z_{\rm top}}$ and $2 |W_{zy}|_{z_{\rm bot}}$ ,

\begin{equation}
\label{wzphi}
|W_{zy}|_{\rm avg} \equiv \frac{1}{2}\left(2|W_{zy}|_{z_{\rm top}} + 2|W_{zy}|_{z_{\rm bot}}\right),
\end{equation}

\noindent
where the factors of 2 result from the assumption made in equation~(\ref{wzphi_symm}), and we take the average since it is not guaranteed that the absolute values of the $zy$ stress are the same at the top and bottom of the domain.

With this information in hand, we estimate the accretion rate by assuming a steady state disk and equating the stress to the accretion rate through the angular momentum evolution equation 
\cite[see, e.g.,][]{simon13b},

\begin{equation}
\label{mdot_tot}
\dot{M} = \frac{2\pi\Sigma\cs^2}{\Omega}\left[\alpha + \frac{4}{\sqrt{\pi}}\frac{R}{H}\overline{|W_{zy}|}_{\rm avg}\right],
\end{equation}

\noindent
where here, $R$ is the radial distance of the shearing box. This equation is only approximate since it was derived by calculating the mass accretion rate due to the $xy$ stress alone and the $zy$ stress alone and then adding the two rates.  Furthermore, since the calculation of the wind stress should really be done in a global simulation as discussed above, the contribution of $\overline{|W_{zy}|}_{\rm avg}$ to $\dot{M}$ should be taken with considerable caution.  With these uncertainties in mind, we use Equation~(\ref{mdot_tot}) to calculate an order of magnitude estimate for $\dot{M}$, which we list in Table~\ref{tbl:sims}.  The accretion rates fall within the range $10^{-8}$-$10^{-7} M_{\sun}/{\rm yr}$, in general agreement with observational constraints for typical T~Tauri stars \cite[e.g.,][]{gullbring98,hartmann98a}.  For the specific case of HD~163296, \cite{mendiguta13} calculated a mass accretion rate $\dot{M} \approx 4.5\times10^{-7} M_{\sun}/{\rm yr}$, while also noting that previous results indicate a lower accretion rate ($\dot{M} \sim 10^{-8} M_{\sun}/{\rm yr}$) suggesting that the accretion rate has increased by an order of magnitude over the past 15 years.  Our simulations reproduce these observed rates, and the runs with $\beta_0 = 10^4$ in particular agree with the currently observed accretion rate for HD~163296.  

\begin{figure}
\begin{center}
\includegraphics[width=0.48\textwidth,angle=0]{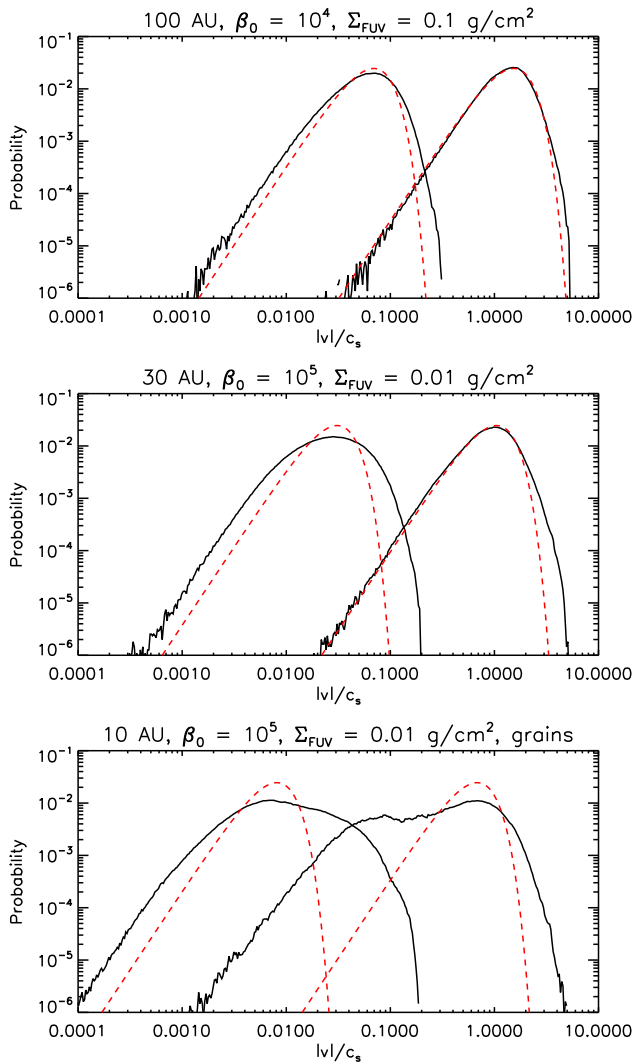}
\end{center}
\caption{
Distribution of turbulent velocities for three representative simulations as labelled on the individual plots.  The turbulent velocities from the simulations are described in the text via Equation~(\ref{rms_velocity}) and are denoted by the black curves.  The red, dashed curves are the best fit Maxwell-Boltzmann distribution. The distributions at the mid-plane (left curves in each plot) and at $z = 4H$ (right curves in each plot) are shown.  Velocities are larger at higher altitudes.
A Maxwell-Boltzmann distribution provides an excellent fit to the simulation data in the top panel and less so in the middle and bottom panels.  The simulation in the bottom panel has the worst fit out of all of our simulations. Most simulations have a fit closer to the top or middle panels. }
\label{bturb}
\end{figure}

\begin{figure*}
\begin{center}
\includegraphics[width=0.8\textwidth,angle=0]{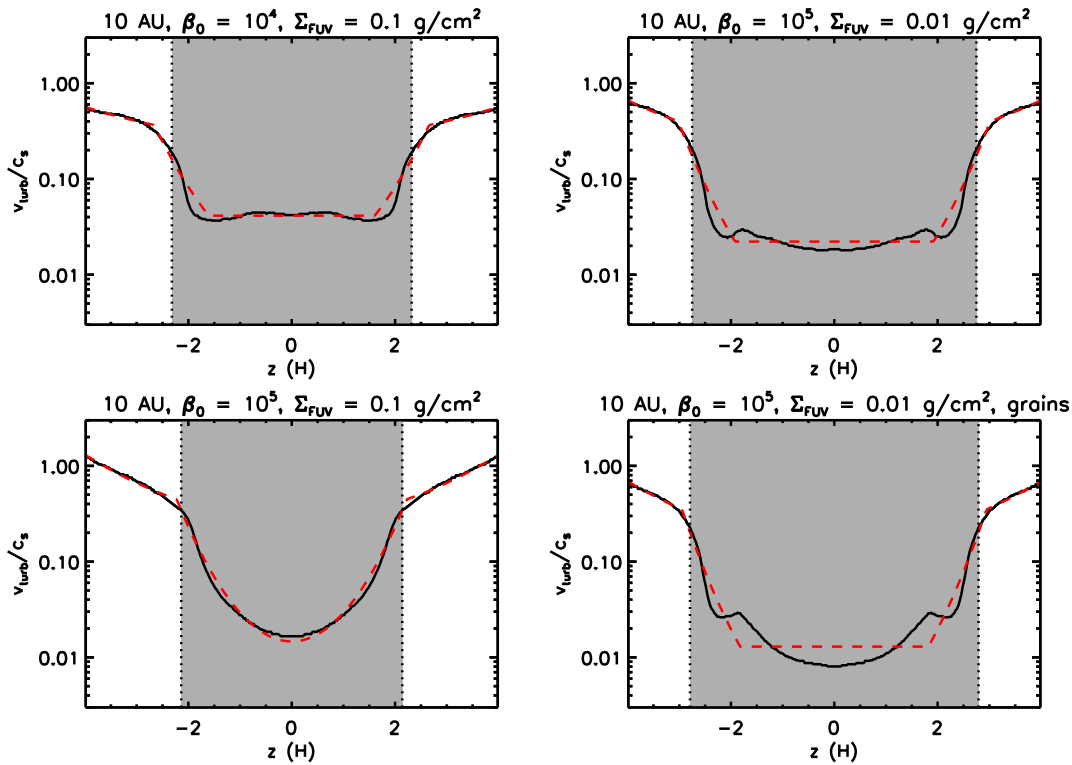}
\end{center}
\caption{
Characteristic turbulent velocity, defined as the peak of a Maxwell-Boltzmann distribution fitted to the velocity distribution, normalized by the sound speed, versus $z$  for the simulations at 10 AU.  The black curves are the data, and the red, dashed curves are analytic fits using the piecewise function defined in Equation~(\ref{vturb_fit}).  The gray regions bounded by dotted lines correspond to the region of high diffusivity (from Ohmic diffusion, ambipolar diffusion, or both) above which FUV photons significantly enhance the ionization fraction. The turbulent velocities increase with $|z|$ in all cases, and this height dependence is steeper at the transition between the diffusion-dominated and FUV-dominated regions.  
}
\label{fit_vp_10au}
\end{figure*}

\begin{figure*}
\begin{center}
\includegraphics[width=0.8\textwidth,angle=0]{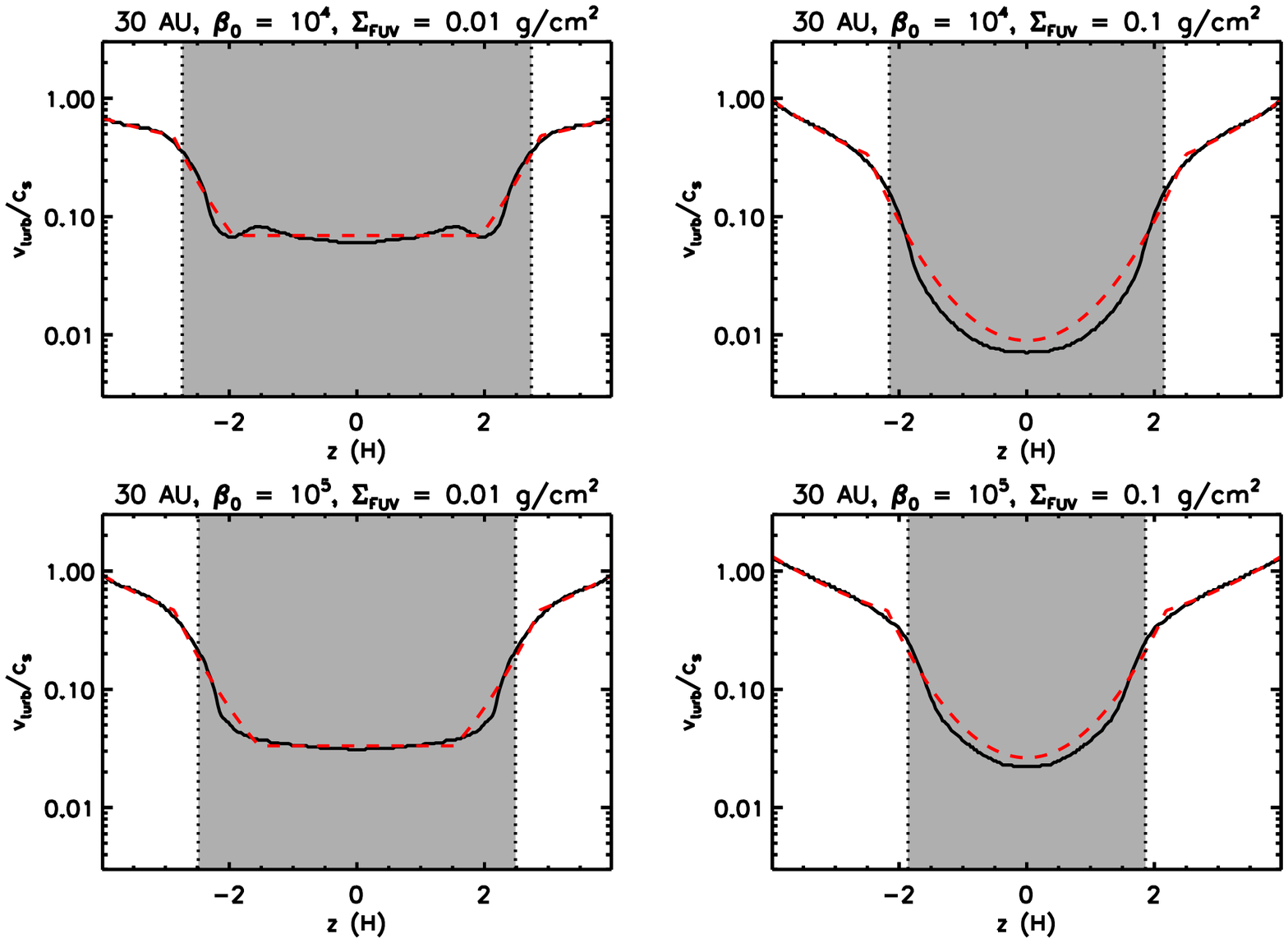}
\end{center}
\caption{
Characteristic turbulent velocity, defined as the peak of a Maxwell-Boltzmann distribution fitted to the velocity distribution, normalized by the sound speed, versus $z$  for the simulations at 30 AU.  We do not include the simulation with grains included because the curve looks nearly identical to the case without grains. The black curves are the data, and the red, dashed curves are analytic fits using the piecewise function defined in Equation~(\ref{vturb_fit}).  The gray regions bounded by dotted lines correspond to the region of high diffusivity (from Ohmic diffusion, ambipolar diffusion, or both) above which FUV photons significantly enhance the ionization fraction. The turbulent velocities increase with $|z|$ in all cases, and this height dependence is steeper at the transition between the diffusion-dominated and FUV-dominated regions.  
}
\label{fit_vp_30au}
\end{figure*}
 
\begin{figure*}
\begin{center}
\includegraphics[width=0.8\textwidth,angle=0]{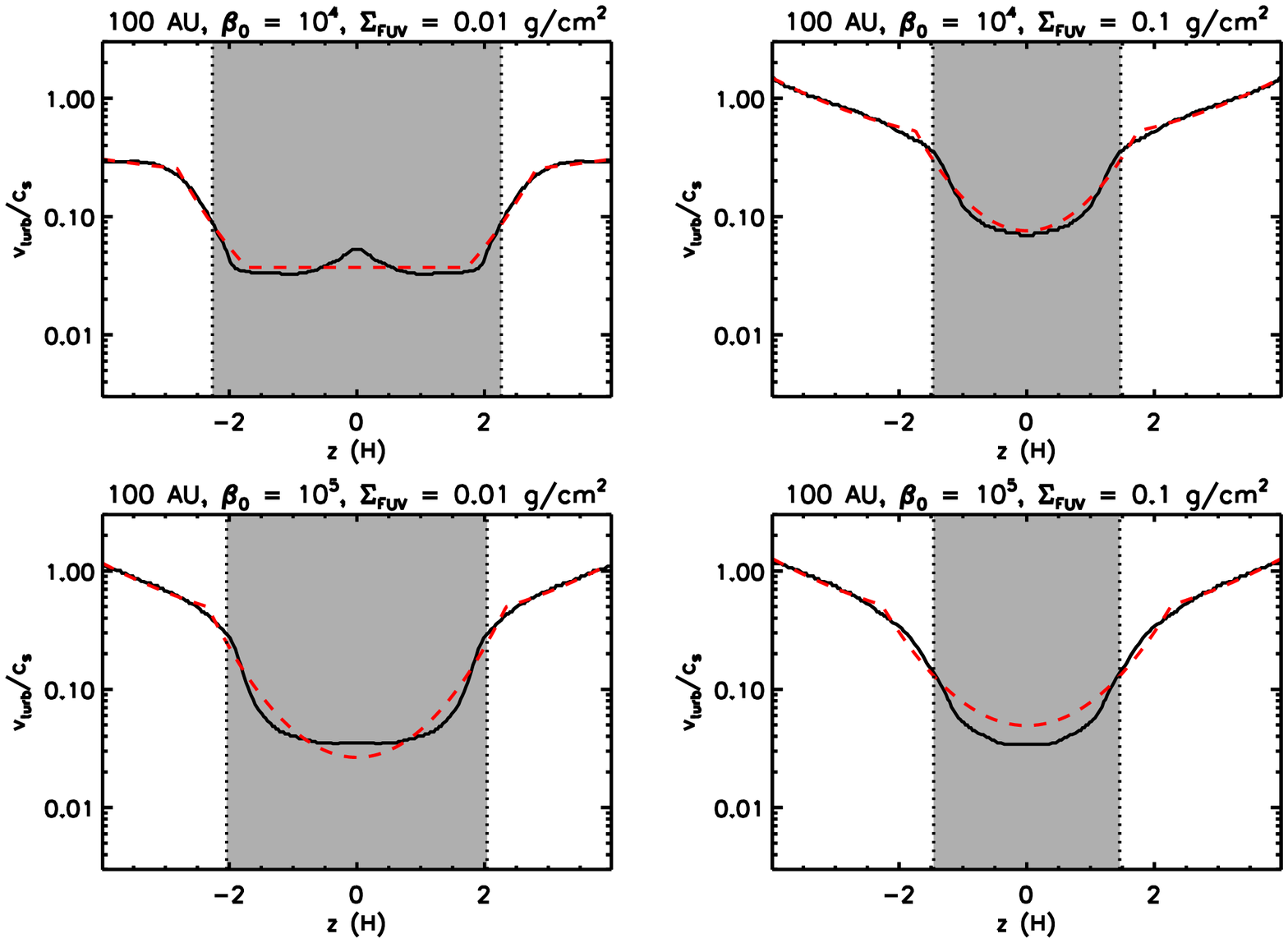}
\end{center}
\caption{
Characteristic turbulent velocity, defined as the peak of a Maxwell-Boltzmann distribution fitted to the velocity distribution, normalized by the sound speed, versus $z$  for the simulations at 100 AU. We do not include the simulation with grains included because the curve looks nearly identical to the case without grains. The black curves are the data, and the red, dashed curves are analytic fits using the piecewise function defined in Equation~(\ref{vturb_fit}).  The gray regions bounded by dotted lines correspond to the region of high diffusivity (from Ohmic diffusion, ambipolar diffusion, or both) above which FUV photons significantly enhance the ionization fraction. The turbulent velocities increase with $|z|$ in all cases, and this height dependence is steeper at the transition between the diffusion-dominated and FUV-dominated regions.  
}
\label{fit_vp_100au}
\end{figure*}

\subsubsection{Turbulent Velocity Distribution}
\label{turb_vel_dist}

We calculate the turbulent velocity distribution from our simulations in order to extract a turbulent broadening parameter for input into LIME.  Because the velocity distribution will change with height above the mid-plane \citep{simon11b} we calculate this distribution as a function of $z$. The total, thermal {\it and} turbulent, broadening can be written as 

\begin{equation}
\label{bturb_tot}
\Delta v(z) =  \sqrt{{2k_{\rm boltz}} T(R,z)/\mu m_{\rm H}+ v_{\rm turb}^2(z)}
\end{equation}

\noindent
where $v_{\rm turb}(z)$ is the characteristic turbulent velocity as a function of height in a given simulation at radius $R$. As we show below, it is generally 
a reasonable approximation to model the small-scale turbulent velocity field with a Maxwell-Boltzmann distribution, maintaining consistency 
between the thermal and turbulent components to the line width. Our basic approach is thus 
to fit Maxwell-Boltzmann distributions to the actual velocity distributions, after appropriate filtering and time-averaging.

To extract the turbulent velocities from our simulations, we must remove any large-scale velocity structures.  In the shearing box approximation with orbital advection the dominant Keplerian motion has already been subtracted off.  However, large scale flows can be still be present.  One example is bulk velocities that fill the entire domain, such as a net radial inflow or outflow.  To remove these, we subtract the horizontally averaged velocities at each height,

\begin{equation}
\label{avg_subtract}
{\bmath v_{\rm red}} = {\bmath v} - \left\langle {\bmath v}\right\rangle_{xy}
\end{equation}

\noindent
where ${\bmath v_{\rm red}}$ is the reduced velocity field and the $xy$ subscript on the angled brackets denote an spatial average in $x$ and $y$.

Another large scale flow often present in shearing box simulations are zonal flows \cite[e.g.,][]{johansen09a,simon12,simon14}, axisymmetric radial variations in the gas density, pressure and azimuthal velocity.\footnote{Zonal flows are a result of geostrophic balance, where pressure gradients are balanced by angular momentum gradients \citep{johansen09a}.}  These flows may well be a physical effect \cite[as suggested by the results of e.g.,][]{bai14c}, but they occur on a radial scale that is large and poorly constrained \citep[indeed they are also seen in global simulations, e.g.][]{dzyurkevich10,flock11,uribe11}. 
If they are observable, it is most likely as a spatially resolved axisymmetric perturbation to the density or velocity field, rather than as a component to unresolved turbulence. Accordingly, we remove them from the velocity field before calculating a velocity distribution.  Since zonal flows show up in the azimuthal velocity and are axisymmetric, we subtract their influence by removing the $y$ average of $v_y$ at each $x$ and $z$,

\begin{equation}
\label{zf_subract}
v_{y,{\rm red}}^{'} = v_{y,{\rm red}} - \langle v_{y,{\rm red}}\rangle_y .
\end{equation} 
Having subtracted the large scale components of the flow, we are still left with a small scale velocity field that varies in space and 
time. To reduce this to a simple scalar function describing the dependence of turbulence on height requires several further steps. 
First, we ignore any anisotropy in the reduced velocity field, and calculate the magnitude of the velocity field 
from which we will calculate a distribution,

\begin{equation}
\label{rms_velocity}
|v| = \sqrt{v_{x,{\rm red}}^2 + (v_{y,{\rm red}}^{'})^2 + v_{z,{\rm red}}^2}.
\end{equation}
Clearly, the (reduced) turbulent velocities in a disk are not guaranteed to be isotropic. However, we have calculated a measure of the anisotropy by dividing the time-averaged $x$ and $z$ kinetic energies by the time-averaged $y$ kinetic energy and taking the square root.  We find that the different velocities differ by at most a factor of three within the time range over which we average.  For our purposes, this is sufficiently close to an isotropic velocity field.

Second, we assume that $|v|$ is statistically in a steady state so that it makes sense to calculate and use the time average of the velocity distribution. By examining the kinetic energy evolution for each simulation, we found that $|v|$ only deviates from a statistically steady state for the simulations at 10 AU, where the kinetic energy can fluctuate by roughly an order of magnitude or more.  This appears to be a result of sudden restructuring of the magnetic field within the domain.  Our time averaging generally starts after initial transients have died away and ends at the end of each simulation.  Since we employ simple arithmetic averaging, our averaged velocity distributions will pick out velocities at the times when the kinetic energy is the largest.  While this clearly represents an uncertainty in our results, we have quantified the effect of this uncertainty on our results by comparing one of our radiative transfer calculations (as described in Section~\ref{predictions}) to a case where we assume zero turbulence for radii at 10 AU or less.   We found that in terms of the integrated molecular line flux, the turbulence levels at radii $\lesssim 10$AU make no difference.  

Third, we examine whether the distribution of $|v|$, after averaging, can be described via a Maxwell-Boltzmann distribution. 
Figure.~\ref{bturb} shows the distribution of $|v|/c_s$ (labelled $|v|/\cs$ in the figure for simplicity) at the mid-plane and at $z = 4H$ for three of our simulations.  Each distribution is  calculated at a number of independent time-slices and then time-averaged.  The black curves are the distributions extracted from the simulations, and the red, dashed curves are a best fit Maxwell-Boltzmann distribution. The top panel shows a simulation that is fit well by a Maxwell-Boltzmann distribution, whereas the simulations in the middle and bottom panels are not fit as well.\footnote{At first glance, it may appear that one could obtain a better fit by changing the width of the distribution.  However, there is only one parameter to fit, $v_{\rm turb}$, which is the peak velocity of the distribution.} 
The Maxwell-Boltzmann fit to the data in the bottom panel is the worst of all of our simulations, but is the only simulation that is fit this poorly; most of our simulations have a fit comparable to what is shown in the top two panels. We believe that this bad fit is a result of poor statistics from large temporal variations in the kinetic energy, as discussed above.  Again, since this simulation is one of the 10 AU calculations, the uncertainties associated with this particular run will not have a significant impact on our results.   

Following the above procedure, we fit a Maxwell-Boltzmann distribution to the extracted turbulent velocity distribution at each height for all of our simulations. The characteristic turbulent velocity $v_{\rm turb}$ is then the peak of the fitted Maxwell-Boltzmann distribution at each height.  This velocity is what will be input as the turbulent broadening parameter in LIME.  The turbulent velocity extracted from this fitting procedure is largely symmetric about the disk mid-plane, but there are small asymmetries.  Since our LIME setup requires a perfectly symmetric broadening parameter, we have symmetrized the velocity profiles by taking the average of $v_{\rm turb}$ for $z < 0$ and $z > 0$ and using this average for both above and below the disk mid-plane.  From this point on, we will use $v_{\rm turb}$ to denote this {\it symmetrized} velocity.

We plot $v_{\rm turb}$ vs $z$ in Figs.~\ref{fit_vp_10au}-\ref{fit_vp_100au} (black curves).  In agreement with previous results \citep{simon11b,fromang06b}, the turbulent velocity generally increases with $|z|$, though in some of our simulations (e.g., R100-B4-FUV0.01-NG; top left panel of Fig.~\ref{fit_vp_100au}), the velocity is relatively constant within the dead/damping zone.  Within the MRI-active regions, this height dependence can be explained to first order by the increase in $\alf$ speed with increasing $|z|$.  Since $\alpha \propto 1/\beta$ in MRI-driven turbulence \cite[e.g.,][]{hawley95a}, $\alpha \sim \va^2 /\cs^2$, and since velocity fluctuations, $\delta v$, are proportional to $\alpha^{1/2} \cs$, we have $\delta v \sim \va$.  Simply put, larger $\alf$ speeds lead to more vigorous MRI turbulence, which enhances turbulent velocities. 

The presence of Ohmic and/or ambipolar diffusion near the mid-plane damps MRI-turbulence, reducing turbulent velocities even more in this region.  As a result, the turbulent velocity profile becomes steeper at the transition between the FUV ionized layer and the diffusion-dominated mid-plane region; this is seen in all of our simulations and is in agreement with the results of \cite{simon11b}, which showed that the turbulent velocity at the mid-plane in simulations with an Ohmic dead zone is significantly smaller than the equivalent in a fully ionized simulation.  We do not fully understand why the turbulent velocity profile is flat in some simulations but more rounded in others, though we suspect that it is related to how turbulent energy is injected into the diffusion-dominated region from the active layers. A more detailed analysis of exactly why the turbulent velocities have the vertical profiles that they do is beyond the scope of this paper, but will be addressed in future work.

As LIME requires an analytic function for $v_{\rm turb}$ vs $z$, we fit the following piecewise function to the curves in Figs.~\ref{fit_vp_10au}-\ref{fit_vp_100au} (red, dashed lines),

\begin{equation}
\label{vturb_fit}
v_{\rm turb, fit}(z) = {\rm MAX}\left[{\rm MIN}\left(a_1 e^{z^2/a_2}, a_3 e^{z^2/a_4} \right), a_5\right],
\end{equation}

\noindent
where the various $a$ coefficients are the free parameters to be fit.  Thus, our fitting function consists of the piecewise combination of two Gaussians and a constant.  The constant $a_5$ is appropriate near the mid-plane for the curves where $v_{\rm turb}$ is relatively flat in that region.  For the other curves, $a_5 = 0$, and this parameter is not even needed in the fitting procedure. As the figures show, the data is reasonably well fit by Equation~(\ref{vturb_fit}).

\section{Observational Predictions}
\label{predictions}

\subsection{Radiative Transfer Calculations}

In order to compare the simulations with observations, we use the Monte Carlo
radiative transfer code LIME \citep{brinch10} in a post-processing
fashion to generate simulated images of the disk projected onto the sky.
In addition to the model parameters describing the disk structure (Table 1),
the simulated image assumes a distance of 122\,pc and inclination to the
line of sight of 44$^\circ$. The assumed density and temperature structures are described in Section 2.1, along with the prescriptions to lower the CO
abundance in the case of freeze-out in the disk interior and photodissociation
on the disk surface.  The velocity field is modeled as a cylindrical Keplerian
field.

The turbulent linewidth is assumed to follow the functional form of Equation~(\ref{vturb_fit}), which 
reproduces the symmetrized velocity derived from the simulations as a function of
height above the midplane to within a factor of $\sim 2$.  We use this
parametric description of the turbulent linewidth primarily for the sake of
computational efficiency.\footnote{LIME is capable of reading an arbitrary tabulated
velocity field but must then perform repeated three-dimensional interpolations
onto the Delaunay grid.}  In addition, the parametric form allows us to smoothly
interpolate between the radii at which the shearing boxes sample (see
description below).  There are three significant assumptions and extrapolations
we must make to graft the results of the shearing box simulations onto a
continuous, two-dimensional velocity distribution in the disk.

First, because the shearing boxes only sample three discrete radii in the disk,
we must interpolate the velocity as a function of radius.  In order to do this,
we perform a simple linear interpolation on each of the parameters $a_1$
through $a_5$ from Equation~(\ref{vturb_fit}).  By interpolating the parameters as a function
of radius, we effectively interpolate between the vertical velocity profiles
displayed in Figs.~\ref{fit_vp_10au}-\ref{fit_vp_100au} so that the turbulent linewidth is defined at all radii
and heights above the midplane throughout the disk.  Interior to the 10\,AU
radius of the innermost shearing box, we make the simplifying assumption that
the vertical profile of turbulence is constant as a function of radius.  This
is unlikely to be true in reality, since radii between 1 and 10\,AU are likely
to be subject to an MRI-inactive ``dead zone" \cite[e.g.,][]{gammie96,bai13b,bai13c}, but for the purposes of this initial investigation, in
which we simulate near-future ALMA observations, these radii are likely to
be too small to be spatially resolved and the assumption will have only minimal
if any effect on the high-velocity wings of the line profile.

Second, there is ample evidence for vertical temperature gradients in
protoplanetary disks, including spatially resolved ALMA observations of the
disk around HD 163296, on which we base our model \citep{rosenfeld13}.  
Our shearing boxes, on the other hand, are vertically isothermal. There is no 
unique way to reconcile these differing structures. Here, we assume that 
the functional form of $v_{\rm turb}(z) / c_s$, derived from isothermal runs, 
holds also when $c_s$ is itself a function of height. 
We therefore calculate the local sound speed of the LIME 
model as a function of both radius and height above the midplane using the
vertically varying temperature structure described in Section 2.1.  
At a given radius, we then use the symmetrized function describing the
turbulent linewidth in units of the sound speed as a function of height above
the midplane to scale the turbulent linewidth to the local sound speed.

Finally, we need to consider what happens when the surface density of the
disk drops below $\Sigma_\mathrm{FUV}$.  For the turbulent simulations, the scaling factor $a_2$
effectively describes the height above the mid-plane at which the vertical
turbulent velocity profile transitions from the higher-velocity Gaussian of
the more strongly ionized upper disk layers to the lower-velocity Gaussian of
the weakly ionized disk interior (see vertical lines in Fig.~\ref{fit_vp_100au} and the arguments in Section~\ref{preliminaries} as to why
radii larger than 100 AU will be in the dynamo state).
The vertical extent of the lower-velocity Gaussian
should gradually taper to zero with increasing radius as the disk surface
density decreases to the value of $\Sigma_\mathrm{FUV}$, at which point the
full vertical column of the disk will be maximally ionized and better described
by the high-velocity Gaussian. To approximate this effect, we gradually taper
the value of $a_2$ beyond 100\,AU so that it reaches a very small value (0.01,
since a value of zero would cause the function to go to infinity) at the
radius at which the surface density equals $\Sigma_\mathrm{FUV}$
(459\,AU for $\Sigma_\mathrm{FUV}$ = 0.01\,g\,cm$^{-2}$ and 308\,AU for
$\Sigma_\mathrm{FUV}$ = 0.1\,g\,cm$^{-2}$).  Beyond this radius, the vertical
profile of turbulence (parameterized as a fraction of the local sound speed) is assumed to be constant with radius.

\subsection{Signatures of Turbulence}

In this section we perform synthetic observations of the turbulent disk model and examine the signatures of turbulence in the disk.  In Section 4.2.1 we examine the spatial and spectral signatures of turbulence in the disk, both qualitatively and quantitatively, within the context of our fiducial model. In Section~4.2.2, we discuss the utility of observing multiple molecular tracers to characterize the vertical profile of turbulence in the disk, and in Section~\ref{sec:resolution} we examine the effects of angular resolution on the strength of the signatures of turbulence for the fiducial model.  Throughout, we base our model on the properties of the bright and well-characterized disk around HD 163296, assuming an inclination to the line of sight of 44$^\circ$.  The observability of turbulence for any given set of observational and source characteristics will naturally depend on a host of parameters specific to the situation, including the specific temperature and density structure of the disk being observed, the distance to and viewing geometry of the system (particularly inclination), the angular resolution, spectral resolution, and sensitivity of the observations, and the properties of turbulence in the disk.  Here we illustrate qualitative and quantitative results from a single representative choice of disk parameters, for the five turbulence simulations presented in the paper, considered against typical observational parameters for a few-hour ALMA data set.  

\begin{figure}[ht!]
\centering
\includegraphics[width=0.45\textwidth]{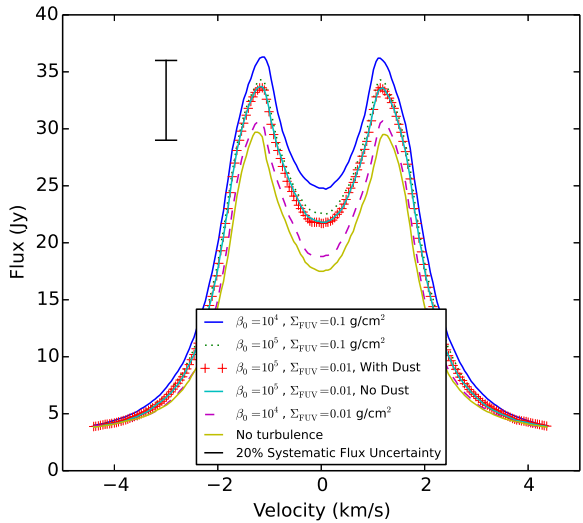}
\caption{Synthetic observations of the CO(3-2) spectra corresponding to a standard density and temperature model of the disk around HD 163296, with the turbulent velocity structure predicted by the various turbulent model conditions described in Section~\ref{setup}.  
Greater turbulent velocities result in a brighter and broader line profile.  
The different model parameters are described in the legend. In the upper left, the scale bar indicates the magnitude of the 20\% systematic flux uncertainty that is currently the best conservative estimate for the accuracy of the absolute flux calibration with ALMA, which is comparable to the predicted difference in flux between the various model cases.  The relative flux uncertainty between channels is far smaller, and depends on the signal-to-noise ratio achieved by the combination of sensitivity and angular resolution in any particular data set.}  
\label{fig:spec_fig}
\end{figure}

\begin{figure}[ht!]
\centering
\includegraphics[width=0.45\textwidth]{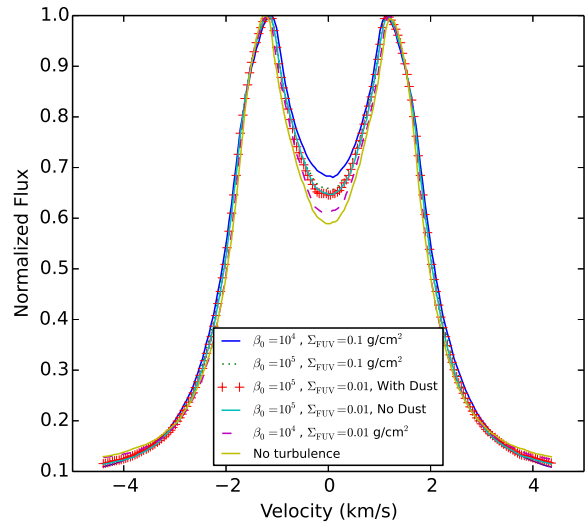}
\caption{
Same as Fig.~\ref{fig:spec_fig}, but in this case the spectra have been normalized to demonstrate how turbulence changes the relative shape of the line.  The systematic flux uncertainty does not affect determination of the relative shape of the line, which is instead determined by the angular resolution and sensitivity of the data.  Diagnostics like the peak-to-trough ratio, which compares the peak flux of the line to the flux at line center, are useful for determining the amount of turbulence present in the protoplanetary disk. 
}  
\label{fig:spec_fig_norm}
\end{figure}

\subsubsection{Observational Signatures of Turbulence in CO~(3-2)}
\label{obs_signatures}

We begin by generating synthetic spectra for each of the different $\beta_0$ and $\Sigma_{\rm FUV}$ conditions described in Section~\ref{setup}.  Fig.~\ref{fig:spec_fig} shows the spectral output of the LIME model of CO J=3-2 emission from the HD 163296 disk, using the fiducial temperature and density structure and imaged at full spectral (44\,m/s) and spatial (10\,AU) resolution, for each of the five different turbulent models.  No noise is added to the spectra, and each channel is summed spatially across the full 12''$\times$12'' simulated sky image.  For comparison, the spectrum of an otherwise identical disk with no turbulent linewidth is also shown.  

In general, the models with greater turbulent velocities yield higher fluxes and broader line peaks.  The spectral broadening is easily understood as the redistribution of flux across velocity channels due to the increased turbulent motion of the medium.  The overall increase in flux is best understood as an optical depth effect: the extra velocity gradient caused by the turbulence effectively spreads some of the flux from each channel into the neighboring velocity channels, and in a highly optically thick medium there is always more CO under the $\tau=1$ surface that would otherwise be obscured and therefore unobservable.  The net effect is that each channel of the line is effectively spatially broadened, and the increase in surface area causes an increase in the flux of the line.  Essentially, CO that would otherwise be invisible in the optically thick disk interior is made visible due to the extra velocity shear caused by the disk turbulence.  This increase in flux is substantial; the model with the largest turbulent velocities exhibits a peak flux $\sim$22\% larger than that of the model with no turbulence.  

Such a dramatic increase in flux may appear easy to detect, but the systematic flux uncertainty of millimeter data is in fact comparable to the expected flux increase, as is illustrated by the scale bar in the figure.  This scale factor ultimately arises from uncertainties in the millimeter flux models of solar system objects, which are typically used to calibrate the absolute flux scale of the data (unlike quasars, which exhibit dramatic flux variations with time, solar system objects are thermal emitters with relatively predictable fluxes).  Currently, the millimeter flux models of solar system objects are assumed to be accurate to within approximately 20\%, although a goal is to improve these models to achieve lower systematic uncertainties in future generations of ALMA data.  The systematic flux uncertainty is an uncertainty on the absolute flux scale of the data, and manifests itself as a scale factor by which the entire spectrum is multiplied.  It therefore afflicts only the absolute value of the flux, rather than the relative uncertainty between different channels in the spectrum.  The channel-to-channel rms noise level is instead set by the sensitivity and angular resolution of the data, which will differ from one data set to the next depending on the integration time, channel spacing, and the baseline lengths in the array.  As these factors are more readily controlled by the observer, measurement of the line shape rather than its overall brightness is a more promising avenue for revealing the presence of turbulence in the disk.  

\begin{figure}[ht!]
\centering
\includegraphics[width=0.52\textwidth]{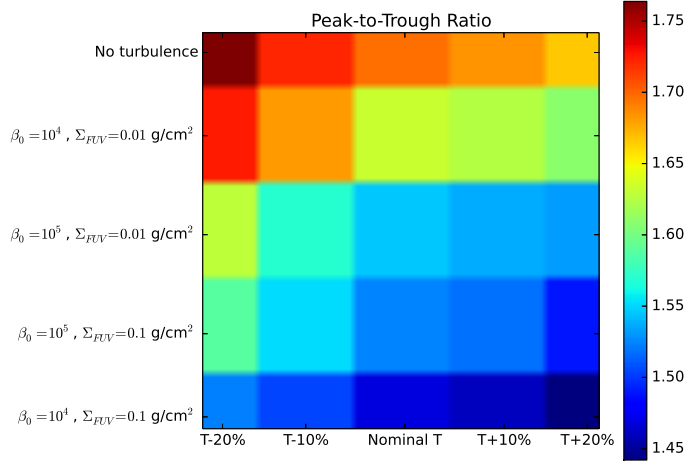}
\caption{
The peak-to-trough ratio, i.e., the ratio of the peak integrated line flux to the integrated flux at line center, for a range of models and assumed temperatures.  The five turbulent model conditions, each of which exhibits a different turbulent linewidth, are arranged along the y-axis in order of most to least turbulent in the CO(3-2) line.  The relative amount of turbulence is inferred from the spectral brightening and broadening evident in Fig.~\ref{fig:spec_fig}, and the ordering of models is the same as in that figure.  Along the x-axis, we vary the temperature structure of the disk relative to the nominal model conditions (in the center of the plot), up to a maximum of $\pm20$\% of the nominal temperature structure.  The range of temperatures is chosen to be consistent with the uncertainty on the temperature structure due to the typical systematic flux uncertainty of 20\% at submillimeter wavelengths.  The peak-to-trough ratio is substantially more sensitive to turbulence than to temperature over the range of temperatures representative of the systematic uncertainty.  
}  
\label{fig:peak_to_trough_fig}
\end{figure}

Fig.~\ref{fig:spec_fig_norm} shows the same model line profiles as in Fig.~\ref{fig:spec_fig}, but normalized by their peak brightness to emphasize the line shape.  It is evident from this figure that the line shape varies substantially with the amount of turbulence in the model, primarily in the degree to which the line center is ``filled in'' by redistributed flux from the peaks: the more turbulence, the more flux is redistributed between channels, and the less flux contrast there is between peaks and center.  The ``peak-to-trough'' ratio is therefore a reasonable diagnostic of the amount of turbulence in the disk, independent of the absolute flux scale.  The peak-to-trough ratio for the most turbulent model is 1.47, compared to a value of 1.69 for the model with no turbulence.  This difference is detectable as long as the relative flux uncertainty, or channel-to-channel noise in the data, is limited to a few percent.  Even in early science operations, ALMA observations routinely achieve this flux uncertainty on bright CO lines in nearby protoplanetary disks in as little as a few hours or minutes, depending on the details of the configuration and spectral resolution of the data.  

In Fig.~\ref{fig:peak_to_trough_fig}, we examine the peak-to-trough ratio as a tracer of turbulence, even in the case of significant uncertainty in the disk temperature.  The systematic flux uncertainty of 20\% induces a temperature uncertainty that is roughly proportional, since in the Rayleigh-Jeans regime the line flux should be approximately linearly proportional to the temperature of the optically thick emitting surface.  We therefore calculate the peak-to-trough ratio (the peak flux of the line divided by the flux at line center) for all of the different models.  The models are arranged along the y-axis of the plot from most turbulent to least turbulent (according to the ordering derived from Figures \ref{fig:spec_fig} and \ref{fig:spec_fig_norm}), while varying the temperature of the nominal disk model by up to $\pm20$\% to simulate the effects of the systematic uncertainty.  The range of peak-to-trough ratios for a given model is small compared to the difference in peak-to-trough ratio for a given assumed underlying temperature distribution and different turbulent conditions.  While there is some small amount of degeneracy, the effects of turbulence dominate over the effects of the assumed temperature structure over the range of temperatures consistent with the typical systematic flux uncertainty of submillimeter data.  

\begin{figure*}[ht!]
\centering
\includegraphics[width=0.9\textwidth]{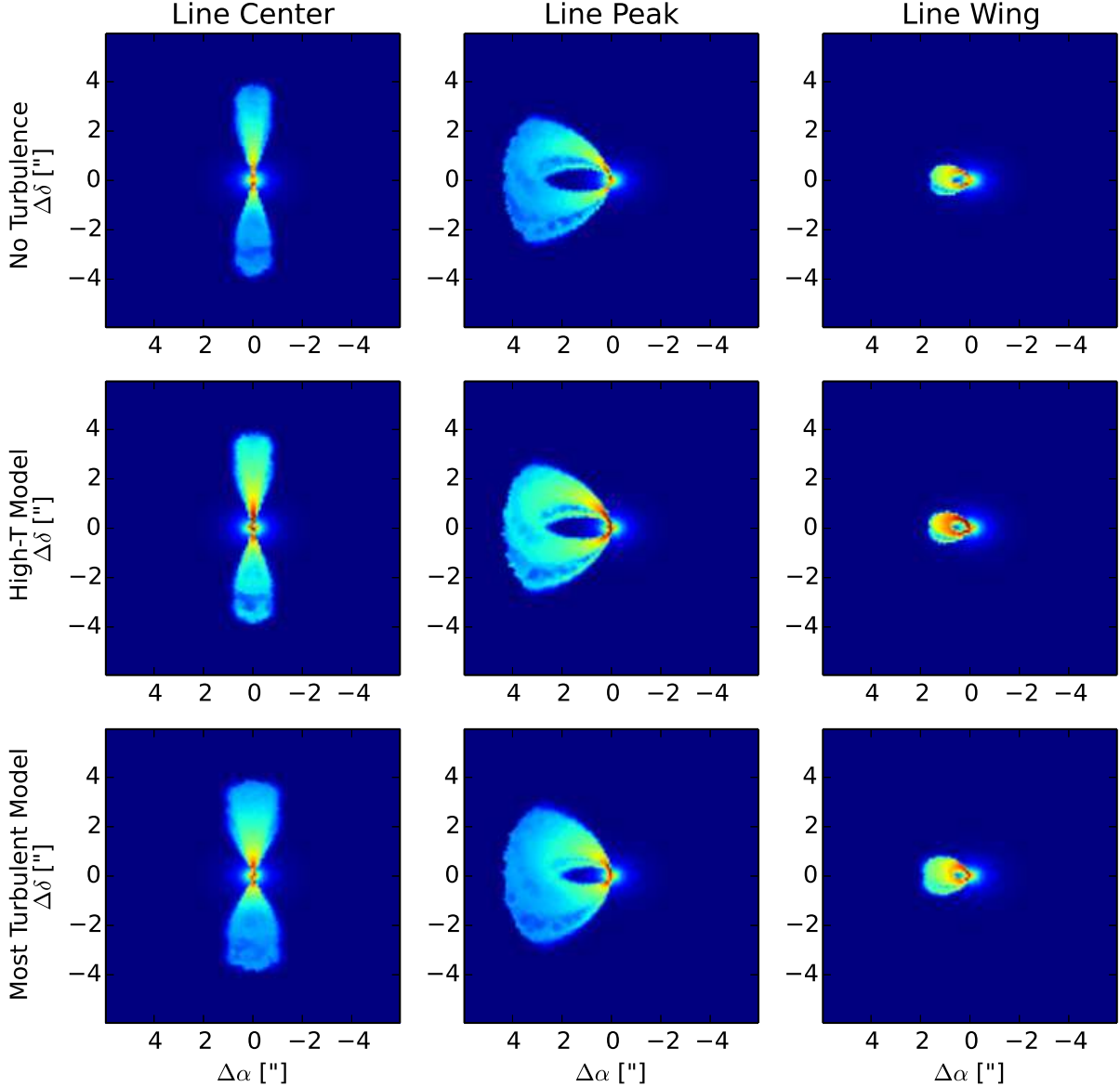}
\caption{
A selection of channel maps generated for simulated images of the turbulent disk models at high (0.08 arcsec) angular resolution.  The left column illustrates the morphology of CO(3-2) emission at line center, i.e., the systematic velocity.  The center column illustrates the morphology near the line peak, and the right column the line wings.  The top row shows the fiducial disk model with no turbulence, the middle row shows the same model with no turbulence but 20\% higher temperature (the maximum allowed by the systematic flux uncertainty), and the bottom row shows the model with the strongest turbulence in the upper disk layers probed by the CO(3-2) line ($\beta_0 = 10^4$, $\Sigma_\mathrm{FUV} = 0.1$~g~cm$^{-2}$).  The primary signature of increased turbulence in channel maps is the spatial {\it broadening} of emission, whereas increased temperature only makes the emission brighter.
}  
\label{fig:channel_map_fig}
\end{figure*}

The effect of increasing the temperature of the disk is spectrally similar to the effect of increasing turbulence (both tend to brighten and broaden the line), however an increase in temperature leads primarily to an increase in surface {\it brightness} of the optically thick disk while an increase in turbulent linewidth manifests itself primarily as an increase in surface {\it area}.  This is illustrated in Fig.~\ref{fig:channel_map_fig}, which demonstrates how the morphology of the channel maps changes as a function of temperature and turbulent linewidth.  The figure displays a selection of channel maps from different regions of the spectrum -- the line center (left), the line peak (center), and the line wings (right) -- for the most turbulent model (bottom row), a model with no turbulence but 20\% higher temperature than the fiducial model (middle row), and a model with no turbulence and fiducial temperature (top row).  Line center refers to the channel corresponding to the systemic velocity, i.e., the channel with no redshift or blueshift along the line of sight relative to the motion of the star.  Line peak refers to velocities near that at which the maximum integrated flux occurs (in the case of our simulated images of HD 163296, these channels occur at roughly $\pm$1.3\,km/s relative to the systemic velocity).  The line wings refer to the channels at velocity extremes, showing the most redshifted and blueshifted parts of the disk relative to the systemic velocity.  For Fig. 10, we have chosen to illustrate channels that are at roughly $\pm$2.6 km/s relative to the systemic velocity.  Each panel of Fig. 13 plots a single 44 m/s wide channel.  Both at line center and in the line wings, the morphology of the emission in each channel is spatially broadened by the presence of turbulence in the disk.  This can be understood as part of the process of redistribution of flux between channels.  The morphology changes with velocity across the line, moving spatially from the peanut-shaped central channels that trace the disk major axis to the ring-shaped channels that bracket the major axis near the line peak.  When turbulence spreads the flux in the spectral domain, it allows emission to ``bleed" between nearby channels, effectively broadening the range of spatial locations that are represented within each channel.  The result is the relatively broader morphology illustrated by Fig.~\ref{fig:channel_map_fig}, in which the models illustrated in the top and bottom rows have  an identical temperature structure but differ in the amount of turbulence that they exhibit.

Interferometric molecular line data is fundamentally three-dimensional in nature, and any fit to the data would naturally take into account the full position-position-velocity cube, including constraints both from the morphology of the line at each velocity and the overall shape of the spectrum.  The simulated images of the model illustrate that both the spectral line shape (independent of the uncertain absolute flux scale) and the morphology of the emission within each channel provide constraints on the amount of turbulence in the disk, and that the turbulent broadening can be distinguished from temperature broadening given sufficient relative flux uncertainty and spatial resolution.  

\begin{figure*}[ht!]
\vspace{-1.4in}
\centering
\includegraphics[width=1.0\textwidth]{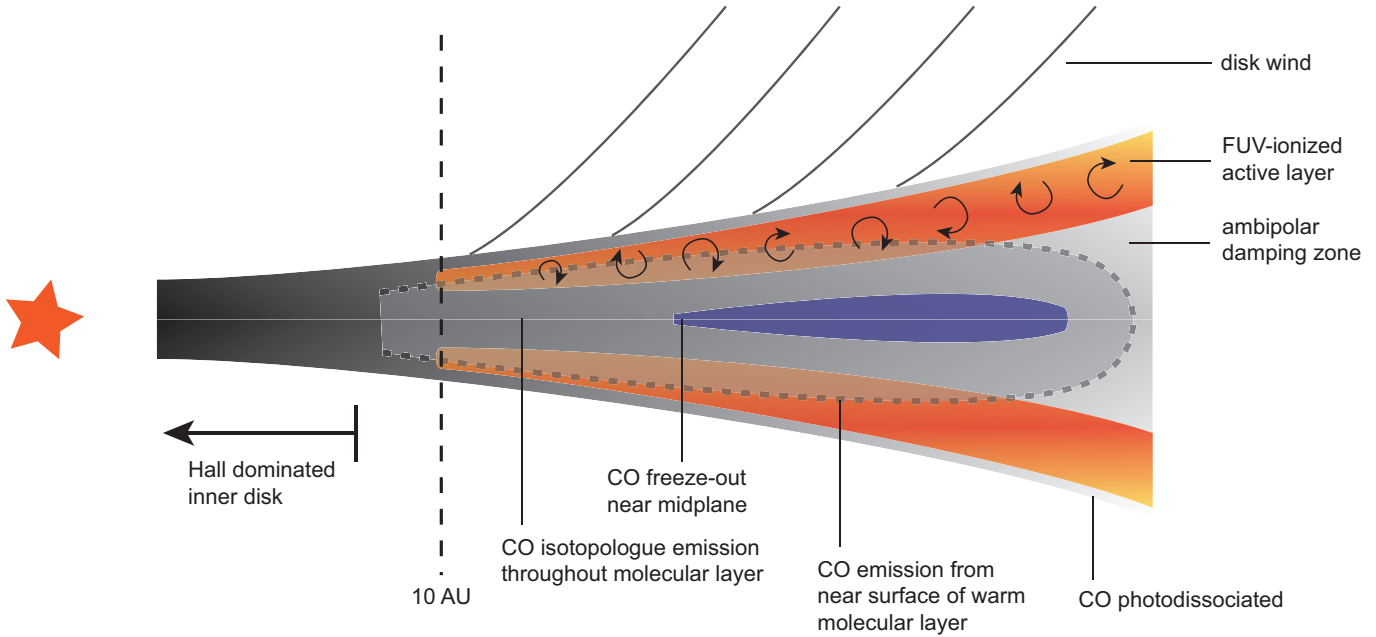}
\caption{
Cartoon showing the layered structure of the disk and illustrating how the optical depth determines the height above the midplane from which different molecular tracers originate.  In the uppermost surface of the disk, CO is photodissociated by energetic stellar radiation and cosmic rays.  In the cold midplane, CO freezes out onto solids and therefore does not emit the gas-phase lines.  CO exists only in the warm molecular layer.  Very optically thick transitions in the CO rotational ladder emit from a region very close to the surface of the warm molecular layer.  CO isotopologues, which have a lower optical depth, have their $\tau = 1$ emitting surface deeper in the disk, where the disk is colder and less turbulent.  In our model of the HD 163296 disk, the densities are high enough that $^{13}$CO is optically thick, while C$^{18}$O remains optically thin throughout the vertical column of the disk.  
}  
\label{fig:disk_layers_cartoon}
\end{figure*}

\subsubsection{The Role of Different Molecular Tracers}
\label{molec_tracers}

Section~4.2.1 illustrated signatures exclusively from the CO(3-2) line, which is one of the most commonly observed tracers of molecular gas in circumstellar disks.  The advantages of observing CO(3-2) are that it is bright and occurs in a relatively optically thin window of Earth's atmosphere; it is therefore a powerful and commonly-observed tracer of the gas component of circumstellar disks.  However, in the context of understanding the turbulent structure of protoplanetary disks, its use is limited.  It is extremely optically thick and probes primarily the upper surface layers of the disk, with the emission originating from a height of 3-5 scale heights above the midplane in our fiducial models.  A judicious choice of molecular line tracers can provide insight into the three-dimensional structure of the disk, allowing the observer to probe much closer to the planet-forming regions near the disk midplane.  

Carbon monoxide is the molecule of choice for most disk observations due to its status as the second most abundant molecule in protoplanetary disks (the most abundant molecule, H$_2$, is virtually invisible due to its lack of a dipole moment).  Observations of CO isotopologues, which have lower optical depths than $^{12}$C$^{16}$O, allow us to probe deeper into the warm molecular layer, as illustrated by the cartoon in Fig.~\ref{fig:disk_layers_cartoon}.  Protoplanetary disks exhibit a vertically and radially stratified structure, including a hot, atomic upper skin in which CO and other molecules are photodissociated; a cold, densely shielded midplane within which volatile species freeze out of the gas phase onto the surfaces of dust grains; and a warm molecular layer in which the conditions are neither too energetic nor too cold for the formation of CO \citep[e.g.][]{aik96,aik06,gor08}.  Efforts at modeling the density structure of disks reveal that in a typical bright protoplanetary disk, both $^{12}$C$^{16}$O (the most common isotopologue, generally abbreviated as just ``CO") and its next-most-abundant isotopologue $^{13}$C$^{16}$O (generally abbreviated ``$^{13}$CO), are optically thick, although $^{13}$CO is more optically thin and therefore originates from a surface that is deeper within the disk \citep[e.g.,][]{pan08}.  $^{12}$C$^{18}$O (generally abbreviated C$^{18}$O) is the next-most-abundant isotopologue, and it is typically the brightest optically thin tracer, dominated by the dense regions close to the midplane but at temperatures too high for CO freeze-out to have occurred.  It is therefore possible to build a vertical profile of turbulence throughout the disk by examining how the peak-to-trough ratio varies for different molecular tracers.  

\begin{figure*}[ht!]
\centering
\includegraphics[width=1.0\textwidth]{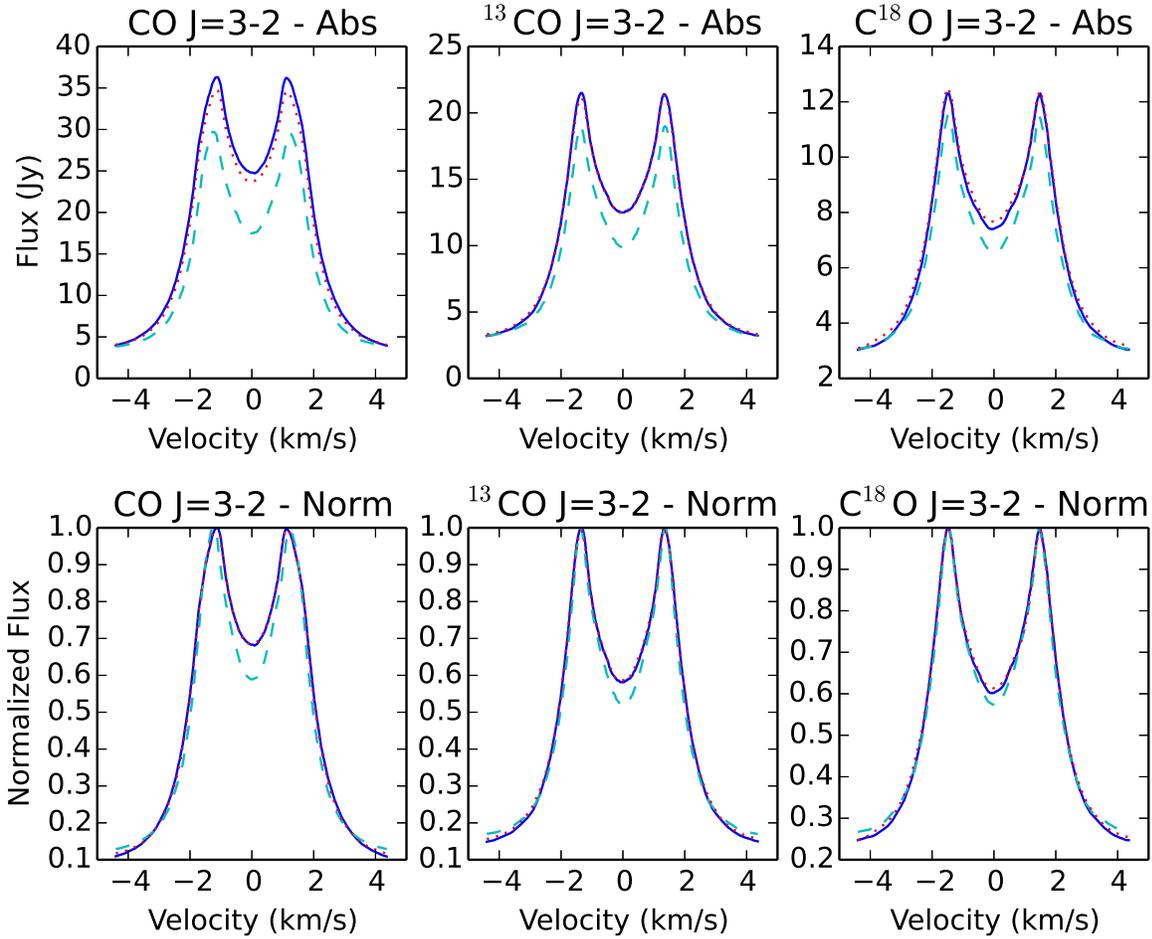}
\caption{Different molecular line tracers with varying optical depth can probe the vertical profile of temperature through the disk.  The top row shows the absolute flux as a function of velocity predicted by the $\beta= 10^4$ $\Sigma_\mathrm{FUV} = 0.1$~g~cm$^{-2}$ model (solid blue line) compared with a model with no turbulence (dashed green line), and a model with a constant turbulent linewidth as a function of sound speed, 0.46\,$c_s$ throughout the disk, chosen to match the normalized CO(3-2) line profile of the most turbulent model (red dotted line).  The bottom row shows the same spectra, normalized to emphasize the peak-to-trough ratio.  The left column shows the predicted CO J=3-2 spectra (highest optical depth), the middle column shows the $^{13}$CO J=3-2 spectra (intermediate optical depth), and the right column shows the C$^{18}$O J=3-2 spectra (lowest optical depth).  The difference in the peak-to-trough ratio for the turbulent vs. non-turbulent models is greatest for the high optical depth tracer that shows the more MRI-active surface of the disk, and smallest for the low optical depth tracer that shows the conditions closer to the midplane.  This reflects the turbulent velocity gradient that increases with distance from the midplane, depicted in Fig.~\ref{bturb}, as well as the effect of the layered disk structure, depicted by the cartoon in Fig.~\ref{fig:disk_layers_cartoon}.  The difference between a constant-turbulence model and a variable-turbulence model is more subtle, especially in the context of the normalized line profiles used to circumvent the ALMA systematic flux uncertainty.}  
\label{fig:iso_fig}
\end{figure*}

This approach is illustrated by Fig.~\ref{fig:iso_fig}.  The top row shows simulated spectra of three isotopologues in terms of the absolute flux, while the bottom row shows normalized flux.  The left column shows CO(3-2), which probes the uppermost surface of the warm molecular layer, while the center shows $^{13}$CO, which probes the middle of the warm molecular layer, and the right column shows C$^{18}$O, which probes the innermost regions of the warm molecular layer close to the CO freeze-out zone in the midplane.  The model with greatest turbulence ($\beta= 10^4$ $\Sigma_\mathrm{FUV} = 0.1$~g~cm$^{-2}$) is represented by the solid blue line, the same model with no turbulence is represented by the dashed green line, and a model with a constant turbulence level of 0.46\,$c_s$, chosen to match the normalized CO(3-2) line profile of the most turbulent model, is represented by the dotted red line.  It is evident that the peak-to-trough ratio increases from the least to most optically thick line tracer, reflecting the increase in turbulent linewidth with height above the midplane.  It is also clear from the figure that the differences in spectral shape between the MHD models in which turbulence varies as a function of height above the midplane, and the constant-turbulence models in which turbulence is a constant fraction of the sound speed (and therefore depends only on temperature), is more subtle, especially in the context of the relatively large ALMA systematic flux uncertainty.  A paper by \citet{hor86} describes the relative lack of sensitivity of optically thin line spectra to levels of turbulence in accretion disks, and emphasizes the importance of the spatial domain, which can provide measurable signatures of turbulence even in the absence of differences in the unresolved line profile.  

While we do not simulate them here, there are other molecular line tracers that are advantageous for probing the three-dimensional structure of turbulence in the disk.  Heavier molecules, like CS, exhibit a naturally lower thermal broadening than CO due to their reduced molecular weight, and therefore turbulent broadening is easier to disentangle from thermal broadening (see Guilloteau et al. 2012).  The disadvantage of such a rare species is that in an optically thin line temperature and density become significantly degenerate, which adds another degeneracy to the turbulent linewidth determination; furthermore, the chemistry of CS and its spatial distribution vertically and radially through the disk is less well understood.  Another promising tracer is DCO$^+$, which is destroyed by CO and therefore becomes abundant in regions where CO is frozen out in the cold midplane.  DCO$^+$ is therefore a promising tracer for placing constraints on turbulent line widths in the midplane of protoplanetary disks.

\subsubsection{The Role of Resolution}
\label{sec:resolution}

Observational characterization of turbulence in disks requires both high angular resolution and high spectral resolution.  The spectral resolution should of course be smaller than the expected turbulent linewidth to ensure a detection (see, e.g., Isella et al. 2007); however, angular resolution is equally important to aid in disentangling the effects of turbulence from those of temperature.  

\begin{figure*}[ht!]
\centering
\vspace{-0.15in}
\includegraphics[width=0.96\textwidth]{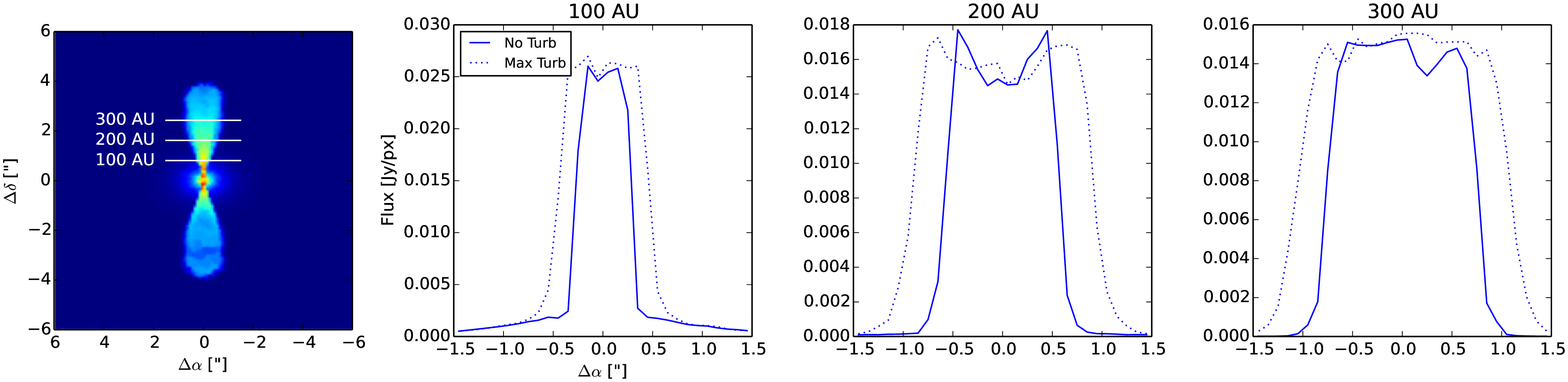}
\includegraphics[width=0.96\textwidth]{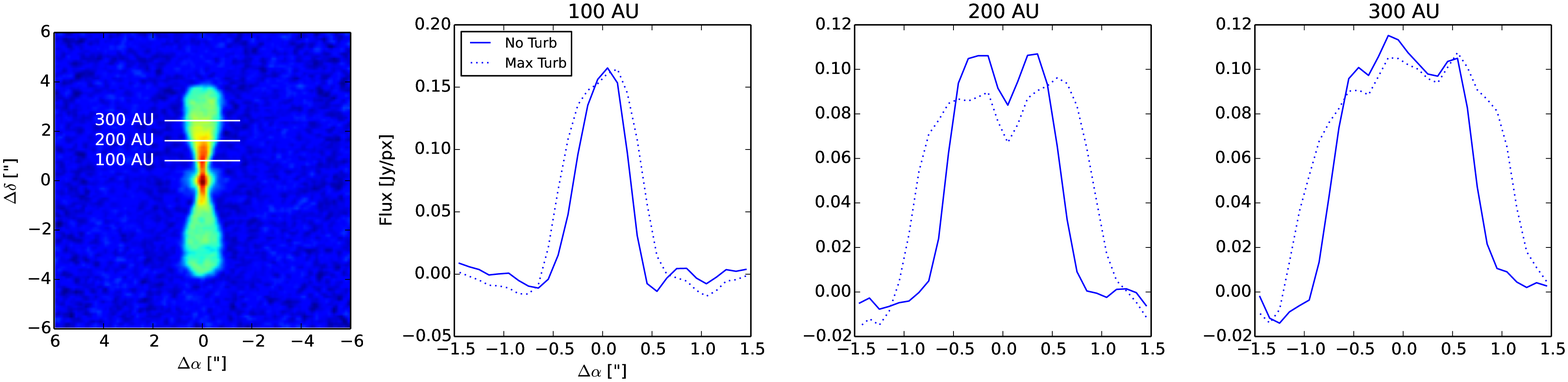}
\caption{A sketch of the role of angular resolution in the detection of turbulence.  The left panel shows the central channel of a model with no turbulence (top left panel of Fig.~\ref{fig:channel_map_fig} above), indicating cuts across the surface brightness of the disk at projected distances of 100, 200, and 300\,AU from the central star.  The three panels to the right show the flux per pixel in the image as a function of offset along the RA cuts indicated in the leftmost panel, at distances of 100, 200, and 300\,AU, respectively.  The non-turbulent model (solid line) is spatially narrower than the turbulent model ($\beta_0 = 10^4, \Sigma_\mathrm{FUV} = 0.1$~g~cm$^{-2}$; dotted line). The top row is for the non-spatially-filtered full-resolution simulated image generated with the Lime radiative transfer code, while the bottom row shows the same models (including appropriate noise) after a three-hour Cycle 3 ALMA observation in the C36-4 configuration at 0\farcs2 angular resolution and 44\,m/s velocity resolution.  The numerical difference in width of the turbulent model relative to the non-turbulent model depends on the chosen spectral channel and the details of the parameters chosen (particularly inclination, spectral resolution, and turbulent linewidth); this figure serves merely to illustrate the importance of spatial resolution in distinguishing turbulent from non-turbulent disk models. 
}  
\label{fig:res_fig}
\end{figure*}

Figure~\ref{fig:res_fig} illustrates the importance of angular resolution in characterizing the turbulent linewidth.  For our fiducial model (based on the observational parameters of HD~163296), we examine three slices in RA across the disk in the central channel (located at the systemic velocity), at projected linear separations of 100, 200, and 300\,AU from the central star.  The locations of the slices are illustrated in the leftmost panel, superimposed over a disk model with no turbulence (taken from the top left panel of Fig.~\ref{fig:channel_map_fig}).  The right three panels show flux as a function of RA offset across each slice.  In each panel, a model with no turbulence (solid line) is compared to the most turbulent model from our simulations ($\beta = 10^4$, $\Sigma_\mathrm{FUV} = 0.1$~g~cm$^{-2}$).  The absolute values of the flux as a function of position are highly dependent on the details of the model (viewing geometry, turbulent linewidth, etc.) and the parameters of the observation (particularly spectral resolution); however the trend of broader emission with greater turbulence is clearly evident.  In the bottom row of the figure we highlight detectability with a realistic simulation of ALMA observations of the model, assuming three hours of observation at 0\farcs2 resolution with the sensitivity anticipated for Cycle 3.  While spatial filtering does somewhat affect the overall flux levels, the trend of increasing spatial width with turbulence is readily evident in realistic simulated observations.  

The different radii show an important trend that emphasizes the necessity of high angular resolution when attempting to characterize turbulence at radii close to the central star.  From the first to the third slice, the distance from the star triples, and the absolute width of the line approximately doubles (as expected for the spatial separation of isovelocity contours of a disk inclined at 44$^\circ$ to the line of sight).  The {\it difference} in spatial width between the turbulent and non-turbulent models over the same range of radii increases more slowly, however, from $\sim$0\farcs4 at 100\,AU to $\sim$0\farcs6 at 300\,AU.  This is also to be expected, since the turbulent linewdith is expected to be proportional to the sound speed, which is proportional to the square root of the temperature, and the temperature decreases roughly as $1/\sqrt{r}$.  For a constant turbulent linewidth as a fraction of the sound speed, therefore, the velocity width of the line should change quite slowly with distance from the star, approximately as $r^{-1/4}$; this will result in a more slowly-changing physical width of the emission as viewed in projection on the sky.  Of course, this $r^{-1/4}$ scaling neglects important properties of turbulence that are included in our models, like the changing ionization fraction and UV penetration depth as a function of radius, and the more shallow radial temperature dependence due to the flaring of the disk.  However, the direction of the trend holds: the turbulent linewidth should change relatively slowly with position in the disk, indicating that the physical width of the line should change relatively slowly with distance from the star.  

This latter trend is good news for studies of turbulent linewidth in the inner disk: as in Fig.~\ref{fig:res_fig}, the total spatial width of the line will decrease rapidly as one moves towards the central star, but the spatial {\it difference} in linewidth between the turbulent and non-turbulent models will fall off more gradually.  Due to the relatively small emitting area of the inner disk, the overall spectral shape of the line emission is dominated by the larger surface area of the outer disk.  High angular resolution is therefore extremely important for detecting turbulence in the inner disk, primarily due to its effects on the spatial width of the line emission.  The exact angular resolution needed to resolve turbulence will depend on the details of the viewing geometry of the target, the expected amount of turbulence in the disk, etc., but in our fiducial highest-turbulence model the spatial turbulent broadening at 100\,AU requires an angular resolution of $\lesssim$0.4\farcs.

\section{Summary and Conclusions}
\label{summary}

Spatially resolved ALMA observations of protoplanetary disks have the potential to yield constraints on turbulence that are more direct than those available in other disk-accreting systems. Maximizing that potential, however -- and determining empirically the nature of protoplanetary disk turbulence -- requires first understanding in detail the observational characteristics of different models of disk turbulence. As a first step toward that goal, we have presented the results of local simulations of magnetohydrodynamic turbulence in the outer disk, incorporating the effects of ambipolar and Ohmic diffusion. The simulations were tailored to match a representative model for the disk around HD~163296, and considered radii 
between 10~AU and 100~AU. We developed and validated a methodology for extracting the turbulent velocity distribution from each simulation, and incorporating it into a radiative transfer calculation that was used to generate synthetic observables.

Our work should be considered as preliminary in several regards. Most obviously, although the numerical simulations include what we believe to be the most essential physics operating on these scales, they have known limitations. The simulations are local, isothermal, neglect the Hall effect\footnote{Preliminary calculations of the turbulent velocity structure that include the Hall effect, by \cite{bai14b}, suggest that this may not substantially impact the results in the outer disk.}, and are limited in terms of the range of parameters studied and length of integration. None of these limitations represent a fundamental obstacle, though relaxing them all (in the absence of algorithmic improvements) would require computing resources well beyond what is currently available. We consider the simplified thermodynamics to be the most serious limitation. 
The assumption of an isothermal disk is not only inconsistent with the radiative transfer calculations but may also affect the propagation of waves through the atmosphere, as wave propagation is known to depend on the vertical thermal structure \citep{bate02}.  

In this analysis we have identified several promising tracers of turbulence and explored in a qualitative way the major known degeneracy between temperature and turbulence.  The observerÕs ability to constrain turbulence given degeneracies with other parameters depends upon the details of the disk and the observational parameters of a given data set, for example, the size, temperature, and viewing geometry of the disk, and the angular resolution and signal-to-noise ratio of the observations.  The question of the observerÕs ability to place strong statistical constraints on the amount of turbulence in the disk given these degeneracies is an important one, but not straightforward to answer in a general way.  A forthcoming publication (Flaherty et al. in prep) will present a thorough statistical analysis of turbulence measurements using ALMA observations of the disk around HD~163296, including an exploration of the degeneracies with related parameters and robust statistical characterization of the degree to which turbulence can be measured given these degeneracies.

Our primary conclusions are as follows:
\begin{enumerate}
\item
Magnetohydrodynamic turbulence in the observable outer regions of the HD~163296 disk is substantially modified by ambipolar diffusion. The MRI can yield accretion rates of the same order of magnitude as those observed if the disk surface is ionized by FUV photons, and a net vertical magnetic field is present. Weaker MRI turbulence and lower accretion rates are possible, but would be inconsistent (within an MRI-dominated disk model) with the idea that the inner disk is sustained from large scale accretion.

\item The characteristic turbulent velocity generally increases with the vertical height above the disk mid-plane, $|z|$.   The slope of this velocity gradient changes at the transition to fully active MRI turbulence in the FUV ionization layer.  In units of the sound speed, this velocity ranges from $v_{\rm turb} \sim 0.01-0.1 \cs$ in the disk mid-plane to  $v_{\rm turb} \sim 0.1-0.4 \cs$ at the bottom edge of the active region.  Within the active region, the turbulent velocity can reach the sound speed, $v_{\rm turb} \sim \cs$.  This height dependence for the turbulent velocity appears to be a generic property of MRI-driven turbulence \citep{fromang06b,simon11b,bai14b}.

\item The outcome of the MRI in the ambipolar dominated outer disk yields two classes of solution, a ``dynamo" solution in which the magnetic field structure in the FUV-ionized layer displays periodic reversals \citep[as in the standard picture of the MRI in ideal MHD, e.g.,][]{simon12}, and a ``non-dynamo" solution in which the magnetic field structure is relatively constant in time and the Maxwell stress has a substantial large scale component.  The dynamo solution appears to be preferred at larger disk radii. The magnitude and vertical gradient of turbulent velocity, however, is essentially independent of the solution type.

\item The simplest observational diagnostic of the level of turbulence is the spatially integrated ratio of the peak flux to the line center flux 
(the ``peak-to-trough ratio"). For the CO(3-2) line, we find readily detectable variations in this ratio between our non-turbulent and most-turbulent models. In contrast to the absolute line flux (which is also a function of the turbulent velocity structure in the disk), the dependence of the peak-to-trough ratio on turbulence is only mildly degenerate with temperature. Moderate temperature uncertainties, resulting from systematic flux uncertainties at millimeter wavelengths, do not prevent a measurement of the turbulent broadening component.

\item Spatially resolved disk observations yield additional diagnostics of disk turbulence. Most obviously, it may be possible to directly observe large scale structures that develop spontaneously within different classes of turbulent flows (e.g. zonal flows in MRI turbulence, or vortices in some alternative models). Small scale 
turbulence manifests spatially as a broader distribution of flux in a given velocity channel (assuming spectral resolution less than the turbulent linewidth), while higher temperature (within the range consistent with a 20\% systematic flux uncertainty) primarily brightens the emission in a given channel.  This effect is most pronounced at the line center and peaks, and has a smaller effect on the line wings.  Due to the relatively shallow dependence of turbulence on radius, the spatial width of the line due to Keplerian velocity shear decreases rapidly, while the spatial width due to turbulence decreases more slowly.  High angular resolution is therefore a potent probe of turbulence in regions close to the central star, whose relatively small emitting areas make spectral probes of turbulence relatively insensitive.  

\item Different molecular line tracers that vary in optical depth trace different heights in the disk. Analysis of an ensemble of line tracers therefore allows us to probe the vertical distribution of turbulent linewidth as a function of height above the midplane. Radiative transfer calculations are needed to determine which lines provide the best discriminants between different turbulence models.

\end{enumerate}

The known properties of the HD~163296 disk, together with existing results from SMA observations \citep{hughes11}, suggest that near-term ALMA observations will be able to constrain protoplanetary disk turbulence through all three of the above diagnostics -- the peak-to-trough ratio, spatially resolved channel maps, and distinct optical depth tracers. Our results suggest that the MRI, at least, displays distinctive observational signatures in each of these diagnostics. Although further theoretical work is needed before precise predictions can be made -- either for the MRI or for other potential drivers of disk turbulence -- the prospects for an observational discrimination of the nature of turbulence in protoplanetary disks with ALMA appear bright.

\acknowledgments

We thank Daniel R. Wik, Matt Kunz, Steve Balbus, Jeremy Goodman, and Richard Nelson for useful discussions and
suggestions regarding this work.  We also thank the anonymous referee, whose suggestions greatly enhanced the quality of
this work. We acknowledge support from NASA through grants NNX13AI58G (P.J.A.) and NNX13AI32G (A.M.H., K.M.F.), from the NSF through grant 
AST 1313021 (P.J.A.), and from grant HST-AR-12814 (P.J.A.) awarded by the Space Telescope Science Institute, which is operated by the  Association of Universities for Research
in Astronomy, Inc., for NASA, under contact NAS 5-26555. J.B.S.'s support was provided in part under contract with
the California Institute of Technology (Caltech) and the Jet Propulsion Laboratory (JPL) funded by NASA through the Sagan Fellowship Program executed by the
NASA Exoplanet Science Institute.  X.N.B. acknowledges support for program number HST-HF-51301.01-A provided by NASA through a Hubble Fellowship grant from the Space
Telescope Science Institute, which is operated by the Association of Universities for Research in Astronomy, Incorporated, under NASA contract
NAS5-26555. This research was supported by an allocation of advanced computing resources
provided by the National Science Foundation. The computations were performed on Kraken at the National
Institute for Computational Sciences and Maverick at the Texas Advanced Computing Center through XSEDE grant TG-AST120062 and on Stampede at the Texas Advanced Computing Center through XSEDE grant TG-AST140001.

\end{document}